\documentclass[]{jfm}

\usepackage{graphicx}
\usepackage{graphics}
\usepackage{newtxtext}
\usepackage[font=small,labelfont=bf,justification=justified,width=\textwidth]{caption}
\usepackage{newtxmath}
\usepackage{natbib}
\usepackage{hyperref}
\usepackage{subcaption}
\hypersetup{
	colorlinks = true,
	urlcolor   = blue,
	citecolor  = black,
}

\newcommand{\RomanNumeralCaps}[1]
\linenumbers

\title{Convective, absolute and global azimuthal magnetorotational
	instabilities}

\author{A. Mishra\aff{1}
 \corresp{\email{a.mishra@hzdr.de}},
 G. Mamatsashvili\aff{1, 2, 3}, V. Galindo\aff{1}
 \and F. Stefani\aff{1}}
\affiliation{\aff{1}Helmholtz-Zentrum Dresden-Rossendorf, Bautzner Landstra\ss e 400, D-01328 Dresden, Germany
\aff{2}Abastumani Astrophysical Observatory, Abastumani 0301, Georgia
\aff{3}Institute of Geophysics, Tbilisi State University, Tbilisi 0193, Georgia}

\begin{document}
\maketitle
	
\begin{abstract}
We study the {\it convective} and {\it absolute} forms of azimuthal magnetorotational instability (AMRI) in a Taylor-Couette (TC) flow with an imposed azimuthal magnetic field. We show that the domain of the convective AMRI is wider than that of the absolute AMRI. Actually, it is the absolute instability which is the most relevant and important for magnetic TC flow experiments. The absolute AMRI, unlike the convective one, stays in the device, displaying a sustained growth that can be experimentally detected. We also study the {\it global} AMRI in a TC flow of finite height using DNS and find that its emerging butterfly-type structure -- a spatio-temporal variation in the form of upward and downward traveling waves -- is in a very good agreement with the linear stability analysis, which indicates the presence of two dominant absolute AMRI modes in the flow giving rise to this global butterfly pattern. 
\end{abstract}



\section{Introduction}
\label{sec:intro}

Many processes of geophysical and astrophysical interest entail the interaction of magnetic fields with conducting fluids or plasma. The dynamo effect is responsible for the self-excitation of magnetic fields in planets, stars and galaxies \citep{Rincon2019}. A key player in cosmic structure formation is the magnetorotational instability \citep[MRI,][]{Balbus_Hawley1998}, which triggers outward transport of angular momentum in accretion disks and mass concentration onto central objects. While dynamo action and MRI are usually considered as separate effects, they are treated as nonlinearly interwoven and mutually reinforcing processes in the modern concept of {\it MRI dynamos} \citep{Rincon2019,Mamatsashvili_etal2020}.

Over the last two decades, a great deal of theoretical and numerical work on dynamo action and MRI has been complemented by a number of pertinent experiments \citep{Stefani_ZAMM2008}. The threshold of self-excitation was achieved in the liquid sodium experiments in Riga \citep{Gailitis_etal2018}, Karlsruhe \citep{Karlsruhe2008} and Cadarache \citep{Cadarache2009}. The helical \citep{Stefani_etal2006} and azimuthal \citep{Seilmayer_etal2014} variants of MRI were observed in the medium-size PROMISE experiment, while an unequivocal proof
of the standard MRI (with an imposed purely axial magnetic field) is still missing, despite great efforts and promising initial results \citep{Sisan_etal2004,Nornberg2010}.

Interestingly, those experimental efforts brought to the fore the distinction between convective, absolute and global instabilities, which are well known in hydrodynamics and plasma physics communities \citep[e.g.,][]{Huerre_Monkewitz1990,Chomaz2005}, but seldom addressed in geophysical and astrophysical magnetohydrodynamics (MHD). It was key for the  success of the Riga experiment to transform the convective instability of the paradigmatic \citet{Ponomarenko_1973} dynamo (a conducting rigid rod, moving in a screw-like manner through an infinitely extended medium with the same conductivity) into an absolute (and a global) one by adding a concentric outer return flow to the inner motion of the rod. The corresponding one-dimensional (1D) eigenvalue problem in the radial direction and the search for saddle points in the complex wavenumber plane, representing the absolute instability, are described in \citet{Gailitis1996}. In this case, the absolute instability for the finite length system coincided quite accurately with the global instability that was later obtained with a two-dimensional code \citep{Stefani_etal1999}. A similar distinction between convective, absolute and global instabilities was discussed for the helical MRI (HMRI) experiment PROMISE \citep{Priede_Gerbeth2009,Stefani_etal2009}. In contrast to the Riga dynamo, where this distinction obviously stems from  the axial flow through the device, the axial propagation of the 
unstable HMRI mode is tied to the direction of the background
Poynting flux \citep{Liu_etal2006}.

This paper is concerned with a related MRI problem for which  the relevance of distinguishing among convective, absolute and global 
instabilities is far less evident. The azimuthal MRI \citep[AMRI,][]{Hollerbach_etal2010} is a non-axisymmetric instability, with dominant azimuthal wavenumbers  $m=\pm1$, arising in a (hydrodynamically stable) differentially rotating flow in the presence of a purely azimuthal magnetic field. While any weak additional axial field would naturally result in a preferred direction and preponderance of either the $m=1$ or $m=-1$ mode, with no axial field these two modes have the same weight and it might be naively expected that they add up to a standing wave. Under this assumption, the two inter-penetrating upward and downward traveling waves, as observed experimentally in \citet{Seilmayer_etal2014}, were interpreted in terms of the breaking of the axial homogeneity by the endcaps of the cylindrical Taylor-Couette (TC) device of PROMISE. However, we will show in this paper that the tendency of AMRI waves to travel away from the mid-height of the cylinders, exhibiting a butterfly-like spatio-temporal variation (analogous to the famous ``butterfly diagram'' in solar physics), has a deeper rooting as it occurs already in a TC flow that is infinitely extended in the axial direction.

\section{Mathematical Formulation}\label{sec:math}

The setup is that of an infinitely long TC flow of a conducting fluid in cylindrical coordinates $(r, \phi, z)$ with an imposed purely azimuthal magnetic field, which is usually adopted in AMRI studies. The inner and outer cylinders with radii $r_{i}$ and $r_{o}$, respectively, rotate with angular velocities $\Omega_{i}$ and $\Omega_{o}$ around $z$-axis, driving in the gap between them an
azimuthal velocity, ${\bf u}_0=r\Omega(r){\bf e}_{\phi}$, with the radial profile of the angular velocity $\Omega(r)= C_1+C_2/r^2$, where  $C_1=(r_o^2\Omega_o-r_i^2\Omega_i)/(r_o^2-r_i^2)$ and $C_2=(\Omega_i-\Omega_o)r_i^2r_o^2/(r_o^2-r_i^2)$. The ratio of the inner to outer radii is $r_i/r_o=0.5$, as in PROMISE. The fluid has constant viscosity $\nu$ and magnetic diffusivity $\eta$ characterized by Reynolds, $Re=\Omega_ir_i^2/\nu$, and magnetic Reynolds, $Rm=\Omega_ir_i^2/\eta$, numbers. The magnetic Prandtl number $Pm=Rm/Re=1.4\times10^{-6}$ is that of GaInSn used in PROMISE \citep{Stefani_etal2009}. In this experiment, an imposed azimuthal magnetic field is current-free, ${\bf B}_0=B_0(r_i/r){\bf e}_{\phi}$, where $B_0$ is constant, which we also adopt here. The effect of the field is quantified by the Hartmann number $Ha=B_0r_i/\sqrt{\rho\mu\nu\eta}$, where $\rho$ is the constant density and $\mu$ is the magnetic permeability. We focus on the Rayleigh stable regime, $\Omega_o/\Omega_i > (r_i/r_o)^2=0.25$, with the fixed ratio of the cylinders' rotation rates $\Omega_o/\Omega_i=0.26$, so that only magnetic instabilities can develop in the flow.

About the above equilibrium TC flow, we consider small perturbations of velocity, ${\bf u}$, total (thermal plus magnetic) pressure, $p$, and magnetic field, ${\bf b}$, which are all functions of the radius $r$ and depend on time $t$, azimuthal $\phi$ and axial $z$ coordinates as a normal mode $\propto {\rm exp}(\gamma t+im\phi+ik_zz)$, where $\gamma$ is the (complex) eigenvalue while $k_z$ and integer $m$ are the axial and azimuthal wavenumbers, respectively. The flow is unstable if the real part (growth rate) of any eigenvalue is positive, i.e., ${\rm Re}(\gamma) > 0$. Henceforth, we normalize length by $r_i$, time by $\Omega_{i}^{-1}$, $\gamma$ and $\Omega(r)$ by $\Omega_i$, ${\bf u}$ by $\Omega_ir_i$, $p$ by $\rho r_i^2\Omega_i^2$, ${\bf B}_0$ by $B_0$ and ${\bf b}$ by $Rm\cdot B_0$. Substituting these non-dimensional quantities in the basic non-ideal MHD equations and linearizing them, we get the perturbation equations also in the non-dimensional form \citep[e.g.,][]{Kirillov_etal2014, Rudiger_etal2018}:
\begin{equation} \label{eq:velocity}
(\gamma + im\Omega){\bf u}+2\Omega {\bf e}_z\times{\bf u}+r\Omega' u_r{\bf e}_{\phi}=-\nabla p+\frac{Ha^2}{Re}\frac{im}{r^2}{\bf b}-\frac{Ha^2}{Re}\frac{2}{r^2}b_{\phi}{\bf e}_r+\frac{1}{Re}\nabla^2{\bf u},
\end{equation}
\begin{equation}\label{eq:mag_field}
Rm(\gamma+im\Omega){\bf b}= \frac{im}{r^2}{\bf u}+Rm\cdot r\Omega'b_r{\bf e}_{\phi}+\frac{2}{r^2}u_r{\bf e}_{\phi} +\nabla^2{\bf b},
\end{equation}
\begin{equation}\label{eq:zerodiv}
 \nabla\cdot{\bf u}=0, \hspace{1cm}  \nabla\cdot{\bf b}=0,
\end{equation}
where $\Omega'=d\Omega/dr$. Equations (\ref{eq:velocity})-(\ref{eq:zerodiv}), together with the adopted no-slip condition for the velocity and perfect-conductor condition for the magnetic field at the cylinder surfaces, constitute the eigenvalue problem, describing AMRI in the magnetized TC flow \citep{Hollerbach_etal2010}. Solving this problem yields the corresponding dispersion relation $\gamma(m,k_z)$ and the radial structure of AMRI modes. Since AMRI is dominated by non-axisymmetric $m=\pm 1$ modes \citep{Hollerbach_etal2010, Rudiger_etal2018}, we focus on these modes in this paper. 

Until now, AMRI has been extensively investigated both with global 1D and local WKB approaches only as a {\it convective instability}  \citep[e.g.,][]{Hollerbach_etal2010, Kirillov_etal2014, Rudiger_etal2018}. These studies formed the basis for understanding the first experimental manifestations of AMRI \citep{Seilmayer_etal2014}. As mentioned above, however, these experiments revealed a notable feature of the propagation of AMRI modes -- the butterfly diagram -- which is not captured by the conventional treatment of AMRI as a convective instability. Our main goal is therefore to show, using the concept of absolute instability, that this butterfly-like propagation is in fact rooted in the dynamics of AMRI itself rather than being induced by the top and bottom boundaries (endcaps) of the TC device.

\section{Convective and absolute instabilities  -- a short overview}

In flow systems, one can distinguish two types of instabilities -- convective and absolute \citep{Huerre_Monkewitz1990,Chomaz2005}. An instability is {\it convective} if the perturbation during growth also propagates in the flow as a traveling wave packet, so that it decays at large times with respect to a fixed point in any reference frame, but continues to grow in the frame co-moving with the group velocity of this packet. By contrast, an instability is {\it absolute} when the perturbation grows unlimited at every point in the flow. In laboratory experiments, it is usually the absolute instability which is more relevant, and hence, of greater interest than the convective instability. An experimental device containing the flow, to which the laboratory frame is attached, has a finite size. As a result, the convective instability can be more rapidly carried out from the system before attaining the growth sufficient for detection, whereas the absolute instability always stays within the system, exhibiting sustained growth detectable in experiments.

The convective instability is determined by a standard procedure of linear modal stability analysis, that is, by considering the dispersion relation at {\it real} wavenumbers $k$ and finding real (growth rate) and imaginary (frequency) parts of $\gamma$ \citep[e.g.,][]{Rudiger_etal2018}. In general, this yields some non-zero group velocity for the unstable modes. Determining the absolute instability is, however, more  delicate, as it requires finding such a combination of convectively unstable modes, which forms growing perturbations (wave packets) with zero group velocity in the laboratory frame, and hence remains in the flow. This is achieved mathematically by considering {\it complex} wavenumbers $k$\footnote{$k$ is the wavenumber in the propagation direction of convective instability, which is along the $z$-axis in the present case.}, analytically extending the dispersion relation $\gamma(k)$ from real $k$-axis to complex $k$-plane and finding its saddle points $k_s$ \citep{Gailitis1980,Huerre_Monkewitz1990,Chomaz2005}, where the complex derivative 
\begin{equation}\label{eq:abs_cond}
\frac{\partial \gamma(k)}{\partial k}\Big|_{k=k_s}=0, \hspace{1.cm} k_s \in \mathbb{C}.
\end{equation}
If the real part of the complex $\gamma$ at $k=k_s$ is positive, ${\rm Re}(\gamma(k_s)) > 0$, this indicates the presence of the absolute instability, where ${\rm Re}(\gamma(k_s))$ is its growth rate and $\omega={\rm Im}(\gamma(k_s))$ its frequency. The real ${\rm Re}(k_s)$ and imaginary ${\rm Im}(k_s)$ parts of $k_s$ describe, respectively, its oscillations and exponential increase/decrease in space. The condition  (\ref{eq:abs_cond}) implies that the real group velocity of the absolute instability is zero at $k_s$, i.e., $\partial \omega/\partial {\rm Re}(k)|_{k=k_s}=0$, although it can still have a non-zero phase velocity.

\section{Convective and absolute types of AMRI}

The AMRI is essentially the instability of inertial waves \citep{Kirillov_etal2014} and therefore it was natural in previous studies to regard it as a convective instability and apply a standard technique of modal linear stability analysis, where the axial wavenumber $k_z$ is real. In this paper, we instead address AMRI within the absolute instability framework, which has not been considered before, and apply the described above methods, where  the wavenumber $k_z$ in the axial $z$-direction, along which AMRI waves travel, is now complex. 

\subsection{WKB analysis} \label{sec:WKB}

To obtain an initial insight into the absolute AMRI, we start with the local short-wavelength WKB analysis, in which the radial dependence of the perturbations is of the form $\exp(ik_rr)$, with a large (real) radial wavenumber $k_rr_i  \gtrsim 1$. Substituting this in (\ref{eq:velocity})-(\ref{eq:zerodiv}) and using the inductionless limit $Pm \rightarrow 0$, we get an analytical dispersion relation \citep{Kirillov_etal2014}:
\begin{equation} \label{eq:disp_rel}
\gamma = -im-\frac{k^2}{Re} - (m^2 + 2\alpha^2)\frac{\textit{Ha}^2}{k^2 \textit{Re}} + 2\left[   \frac{\alpha^2Ha^2}{k^2Re^2}\left((\alpha^2+m^2)\frac{Ha^2}{k^2} +im(Ro+2)\right) - (Ro+1)\alpha^2\right]^{1/2}
\end{equation}
where $\alpha=k_z/k, ~k^2=k_r^2+k_z^2$, $Ro = r(2\Omega)^{-1}d\Omega/dr$ is the Rossby number. Using the conversion formula of \citet{Stefani_Kirillov2015} for $\Omega_o/\Omega_i=0.26$ in the TC flow, we obtain $Ro=-0.97$ in the local case, which is below the lower Liu limit, $Ro_{LLL}=-0.83$, ensuring the presence of AMRI \citep{Kirillov_etal2014}. Typically, AMRI modes extend over the gap width between the cylinders (see below), so for the effective radial wavenumber we take $k_r=\pi/(r_o-r_i)$ \citep{Ji_etal2001}, though this choice is still somewhat arbitrary. 

\begin{figure}
\begin{minipage}{.33\textwidth}
 \includegraphics[width=\textwidth]{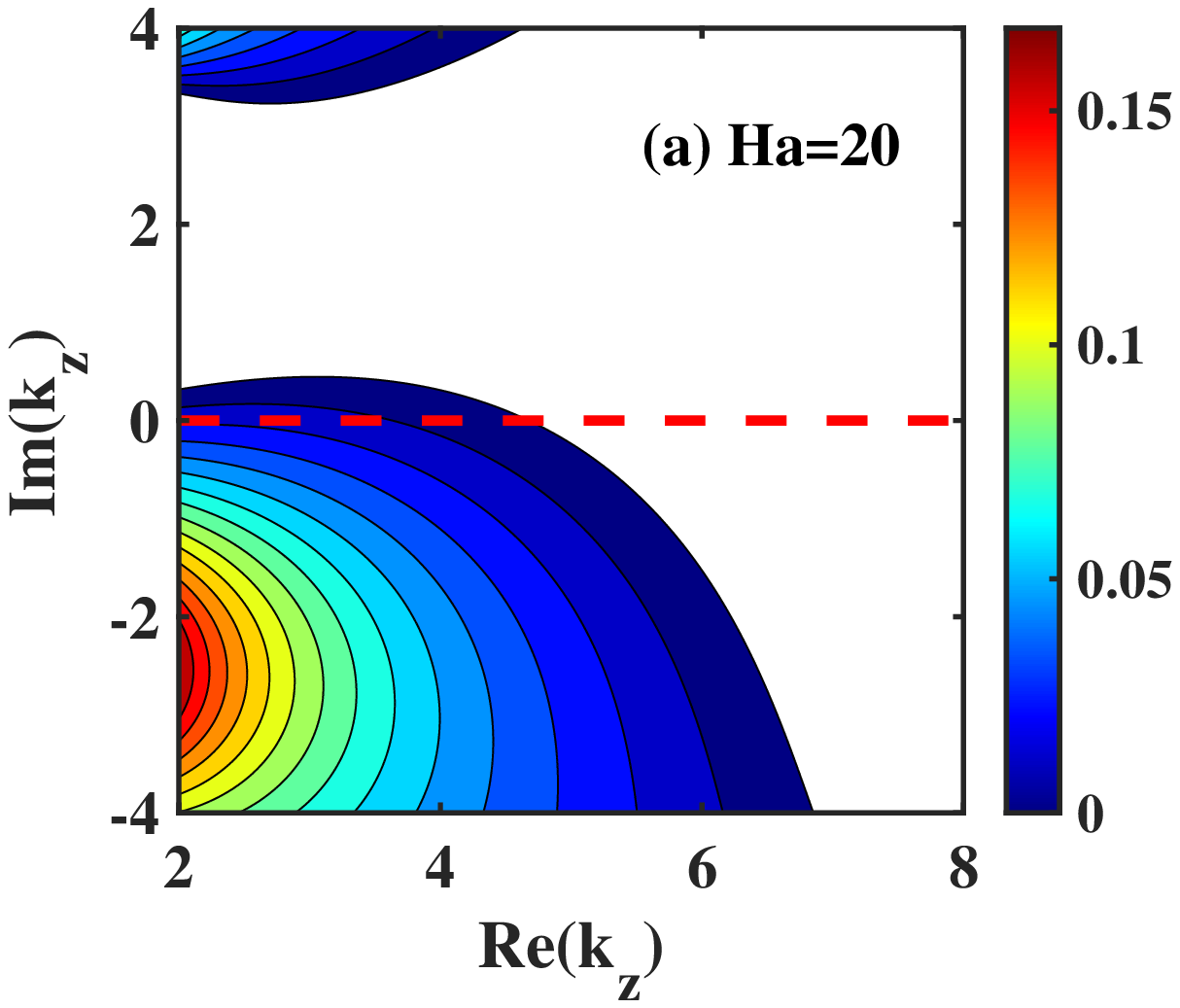}
\end{minipage}
\begin{minipage}{.33\textwidth}
 \includegraphics[width=\textwidth]{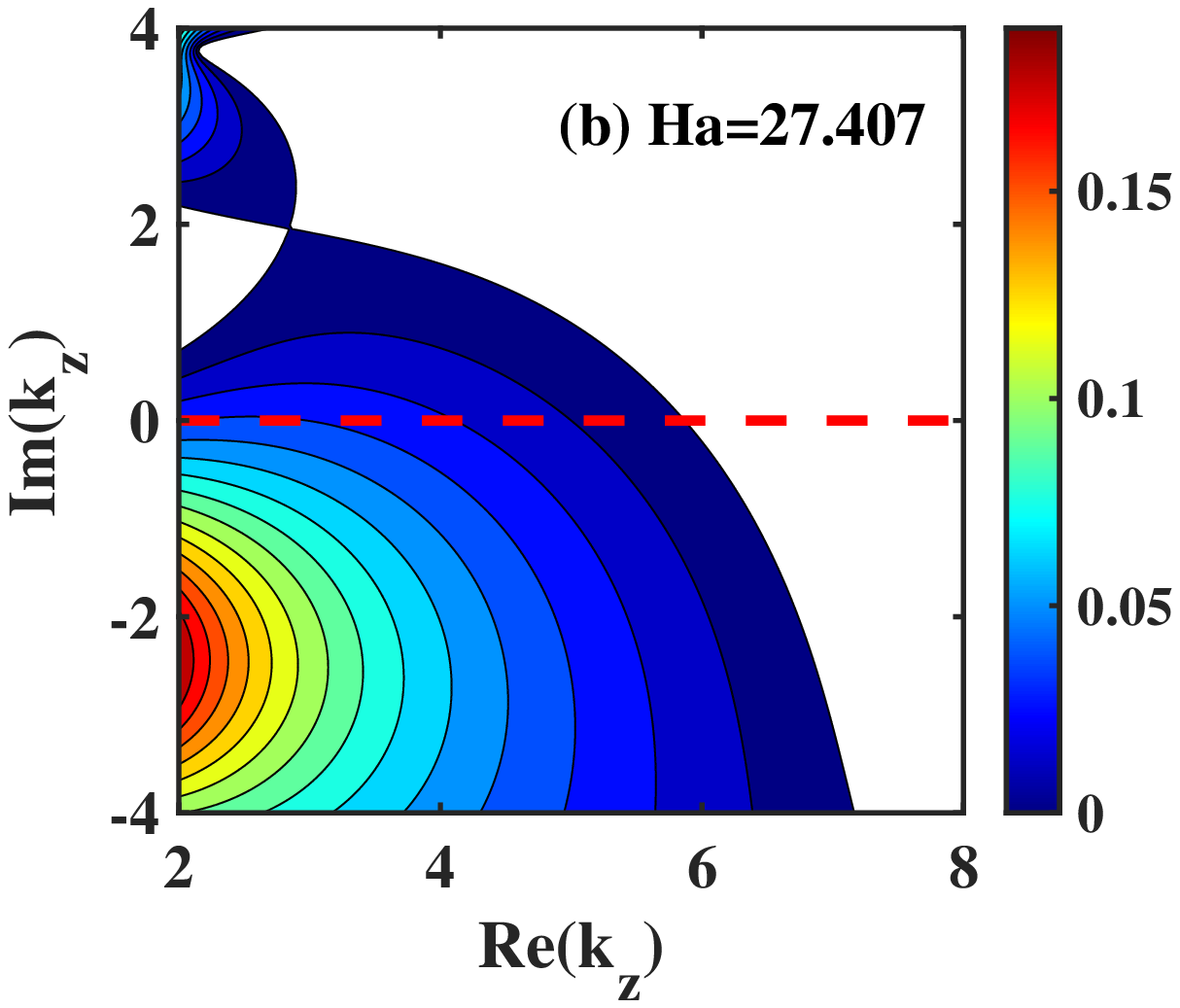}
\end{minipage}
\begin{minipage}{.33\textwidth}
 \includegraphics[width=\textwidth]{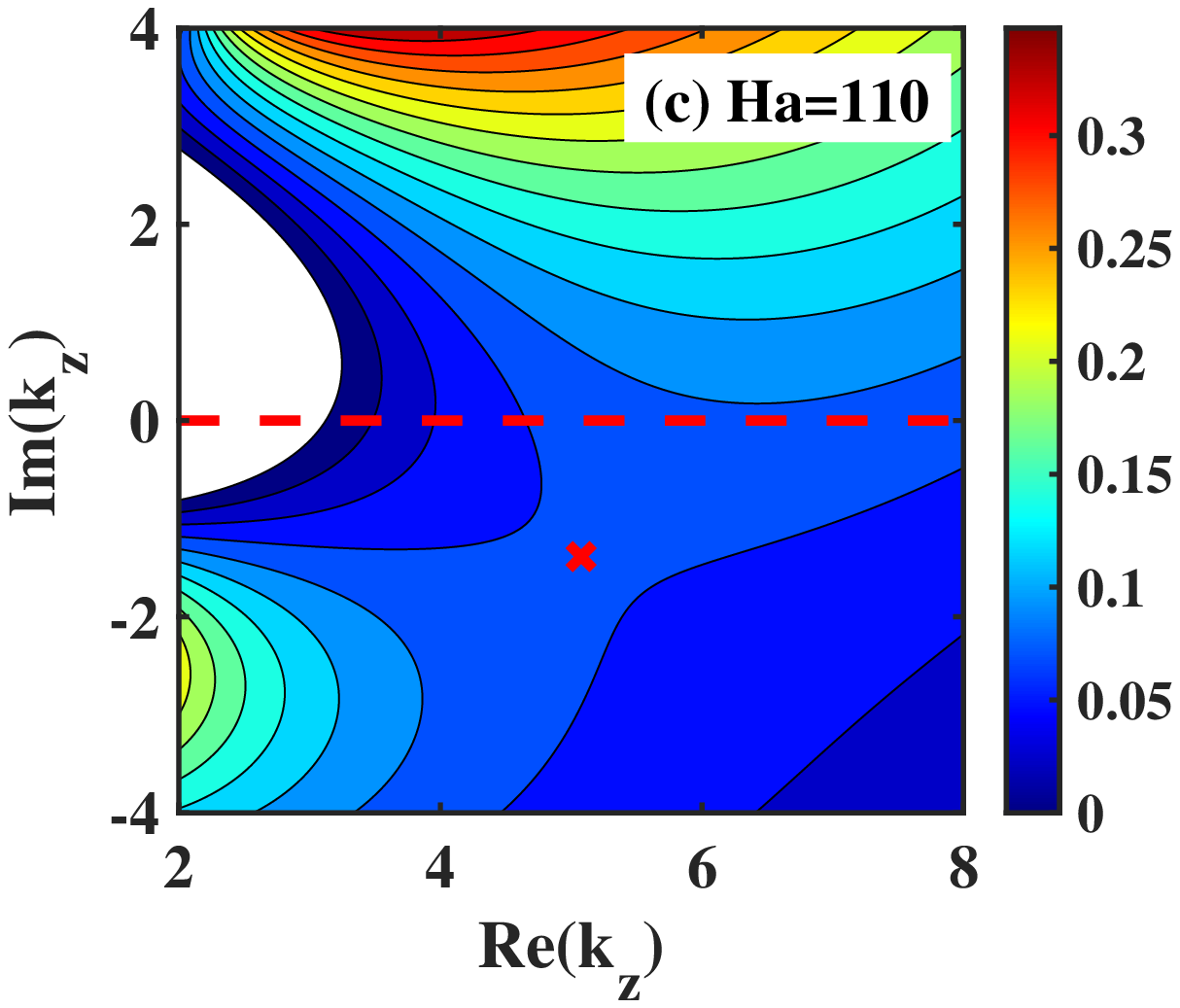}
\end{minipage}
\begin{minipage}{.33\textwidth}
 \includegraphics[width=\textwidth]{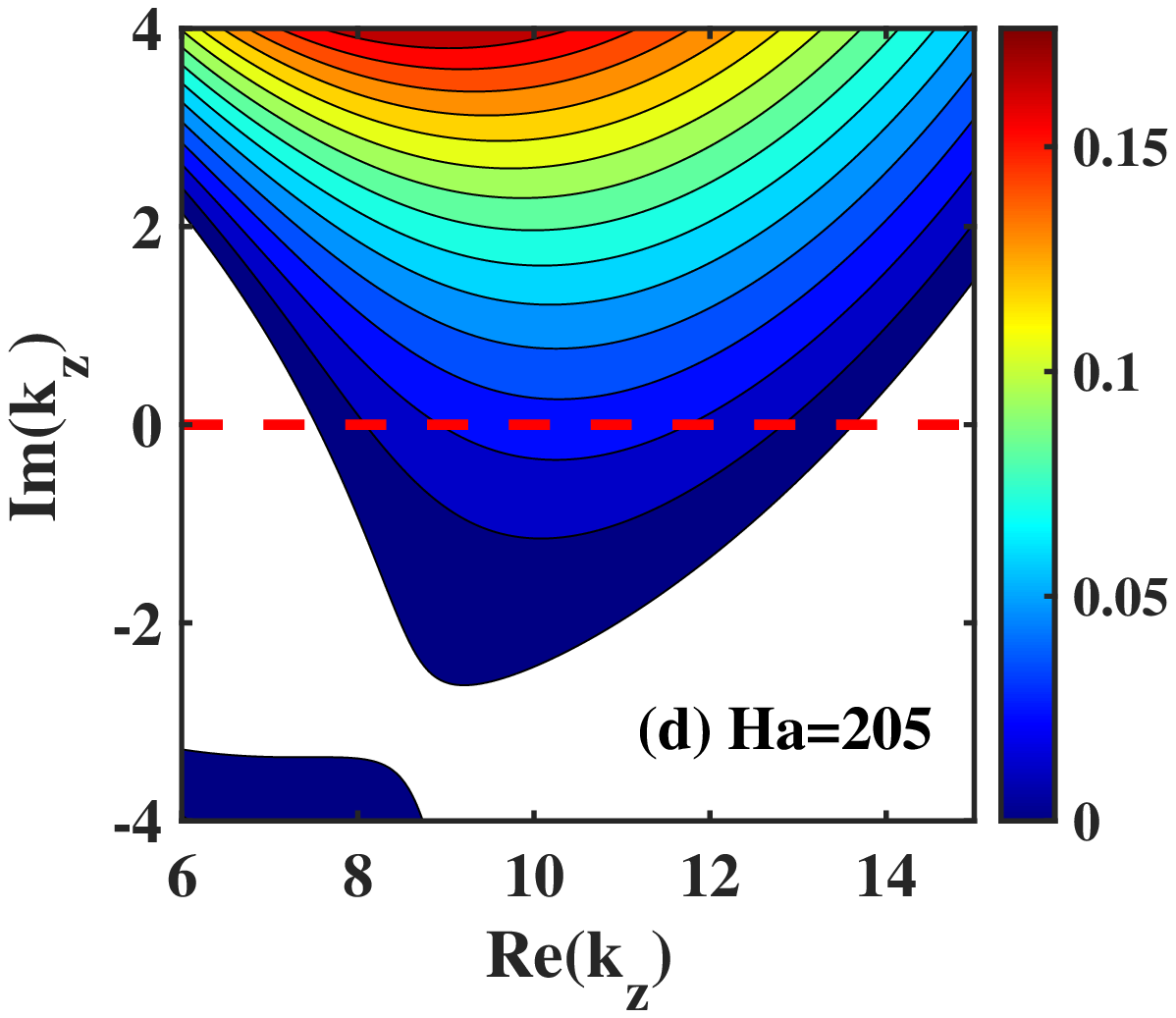}
\end{minipage}
\begin{minipage}{.33\textwidth}
 \includegraphics[width=\textwidth]{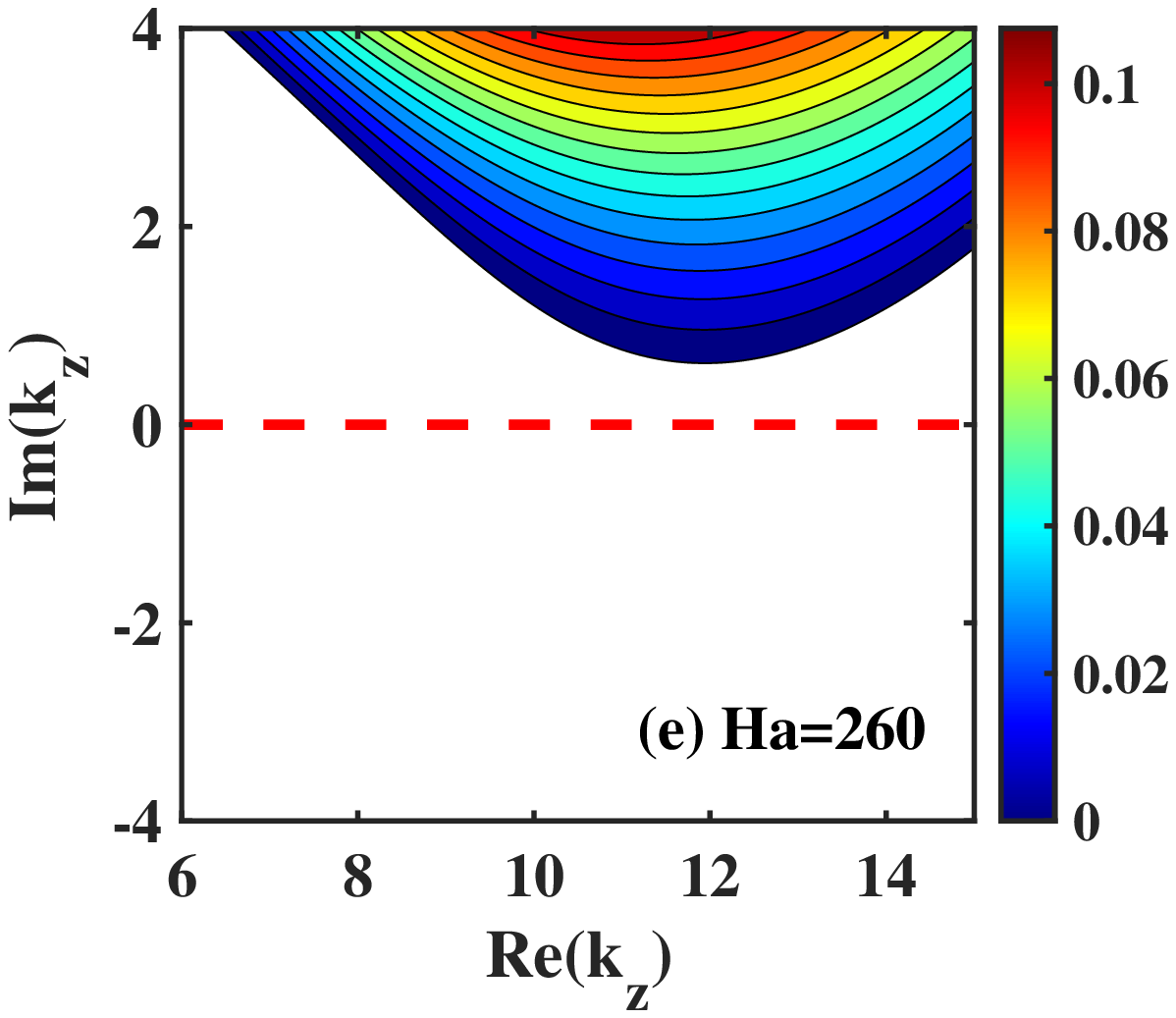}
\end{minipage}
\begin{minipage}{.33\textwidth}
\includegraphics[width=0.875\textwidth]{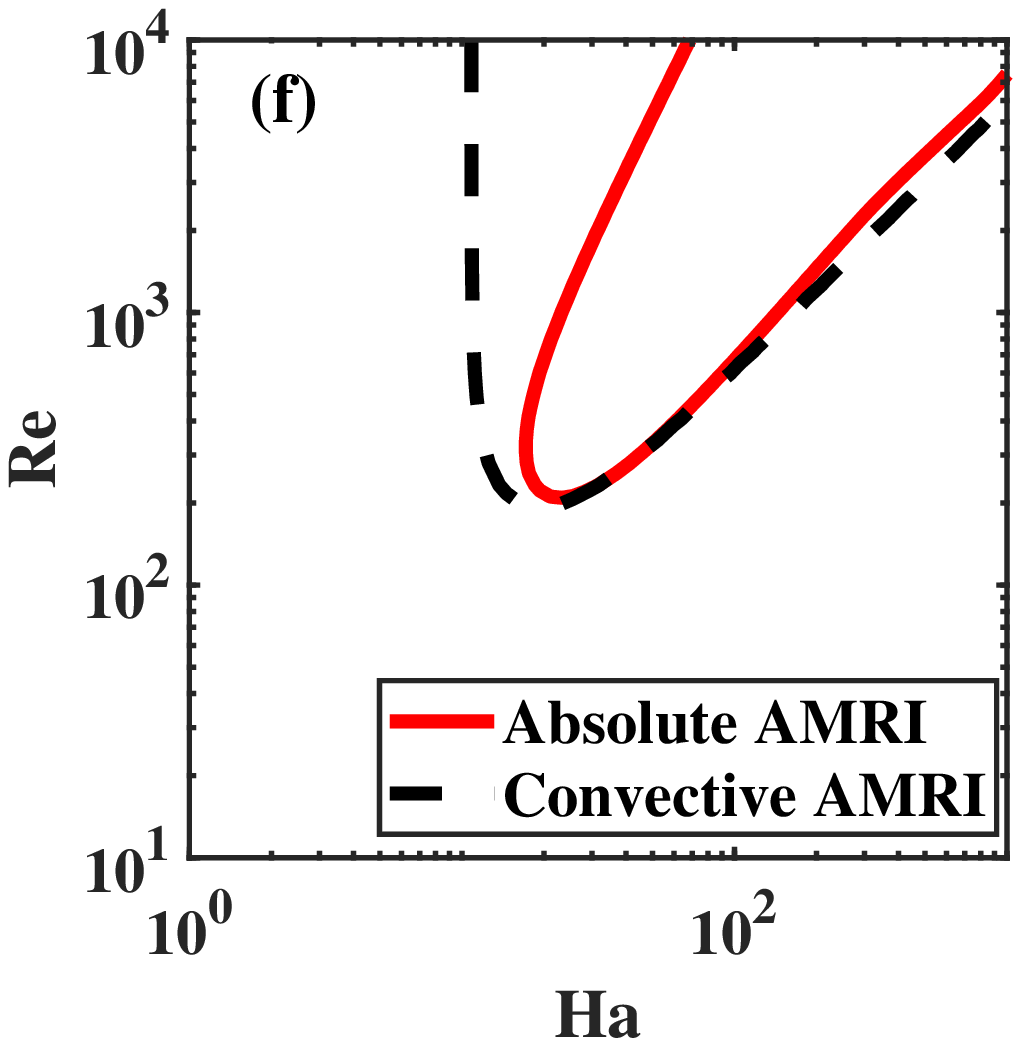}
\end{minipage}
\vspace*{-0.5em}
\caption{(\textit{a})-(\textit{e}) The areas of growth, ${\rm Re}(\gamma)>0$, in the $k_z$-plane obtained from the WKB dispersion relation (\ref{eq:disp_rel}) for fixed $Re=1480, m=1$, $Ro=-0.97$ and different $Ha$. The red cross in (\textit{c}) denotes the saddle point at $k_{z,s}=(5.08, -1.39)$ -- the wavenumber of the absolute AMRI mode with the growth rate ${\rm Re}(\gamma(k_{z,s}))=0.07$. (\textit{f}) The marginal stability curves for the convective (dashed) and absolute (red) AMRI in $(Ha,Re)$-plane.}
\label{fig:WKB}
\end{figure}

Figure \ref{fig:WKB} shows the solution of the dispersion relation (\ref{eq:disp_rel}) for complex $k_z$ at $Re=1480$, $m=1$ and different $Ha$.\footnote{The $m=-1$ case is similar except that the two growth areas swap their ${\rm Im}(k_z)$ values in the $k_z$-plane, so we do not show it here; see Sec. \ref{sec:1D_analysis}.} At smaller $Ha$, there are two separate areas of positive ${\rm Re}(\gamma)\geq 0$, however, 
neither crosses yet the ${\rm Re}(k_z)$-axis (red dashed lines in figure \ref{fig:WKB}), implying that there is no genuine instability in the flow. With increasing $Ha$, the lower area first crosses the ${\rm Re}(k_z)$-axis, indicating the onset of the convective AMRI, and then extends into the upper half of the $k_z$-plane, which is shown at $Ha=20$ in figure  \ref{fig:WKB}(\textit{a}). At $Ha=27.407$, the boundaries (${\rm Re}(\gamma)=0$) of these two areas touch each other, forming a saddle point $k_{z,s}$ according to condition (\ref{eq:abs_cond}), which is seen as a cusp in figure \ref{fig:WKB}(\textit{b}). This saddle point, still being marginally stable (${\rm Re}(\gamma(k_{z,s}))=0$), represents the emerging absolute AMRI. It further develops by increasing its growth rate and spreading towards the lower half of the $k_z$-plane, as the two areas merge more with increasing $Ha$. For example, at $Ha=110$ in figure \ref{fig:WKB}(\textit{c}), the saddle point associated with the absolute AMRI mode is at $k_{z,s}=(5.08, -1.39)$ (red cross). It has the growth rate ${\rm Re}(\gamma(k_{z,s}))=0.07$ and frequency ${\rm Im}(\gamma(k_{z,s}))=-0.5$. At higher $Ha$, the saddle point and hence the absolute AMRI disappears, but the convective instability along the ${\rm Re}(k_z)$-axis can still remain, figure \ref{fig:WKB}(\textit{d}). Eventually, the convective AMRI also vanishes after some maximum $Ha$ is exceeded, as the growth areas move away from the ${\rm Re}(k_z)$-axis, figure \ref{fig:WKB}(\textit{e}).

The marginal stability (${\rm Re}(\gamma)= 0$) curves for convective (optimized over ${\rm Re}(k_z)$) and absolute AMRI are shown in figure \ref{fig:WKB}(\textit{f}), with the latter being located inside the former. This implies that the critical $Ha$ and $Re$ for the excitation of the convective AMRI are typically lower than those for the absolute AMRI. This can be graphically understood from figures \ref{fig:WKB}(\textit{a})--\ref{fig:WKB}(\textit{e}). Either of the two growth areas, approaching from upper and lower halves of the $k_z$-plane, first overlaps ${\rm Re}(k_z)$-axis at certain $Ha$ for a given $Re$, which lies on the black dashed curve in figure \ref{fig:WKB}({\textit{f}}), indicating the onset of the convective AMRI. Then, they touch each other at some larger $Ha$ that lies on the red curve in figure \ref{fig:WKB}({\textit{f}}), indicating the onset of the absolute AMRI. If the saddle point lies on the ${\rm Re}(k_z)$-axis, then the marginal stability curves for convective and absolute AMRI coincide.

\begin{figure}
\begin{minipage}{.33\textwidth}
\centerline{\includegraphics[width=\textwidth]{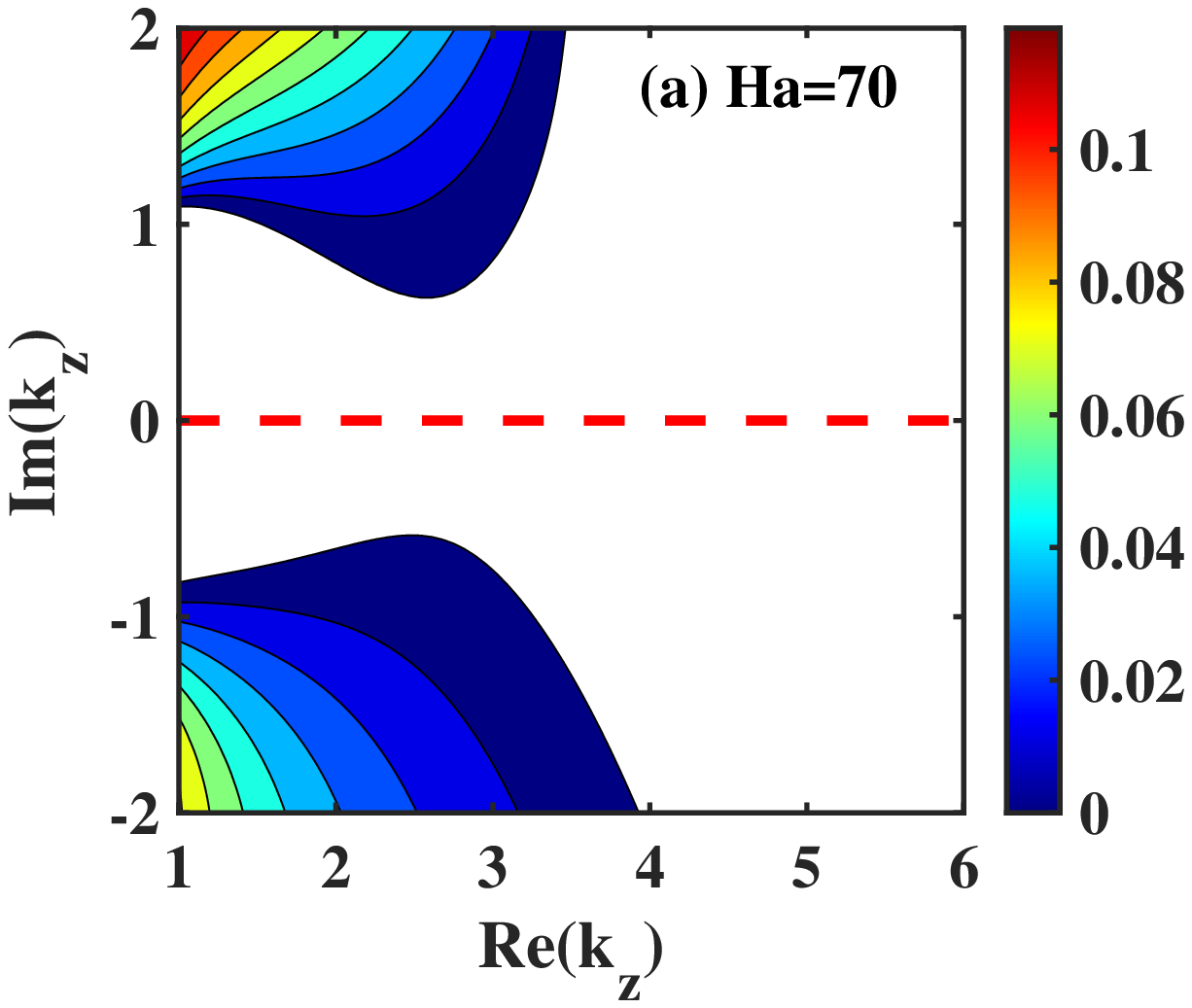}}
\end{minipage}
\begin{minipage}{.33\textwidth}
\centerline{\includegraphics[width=\textwidth]{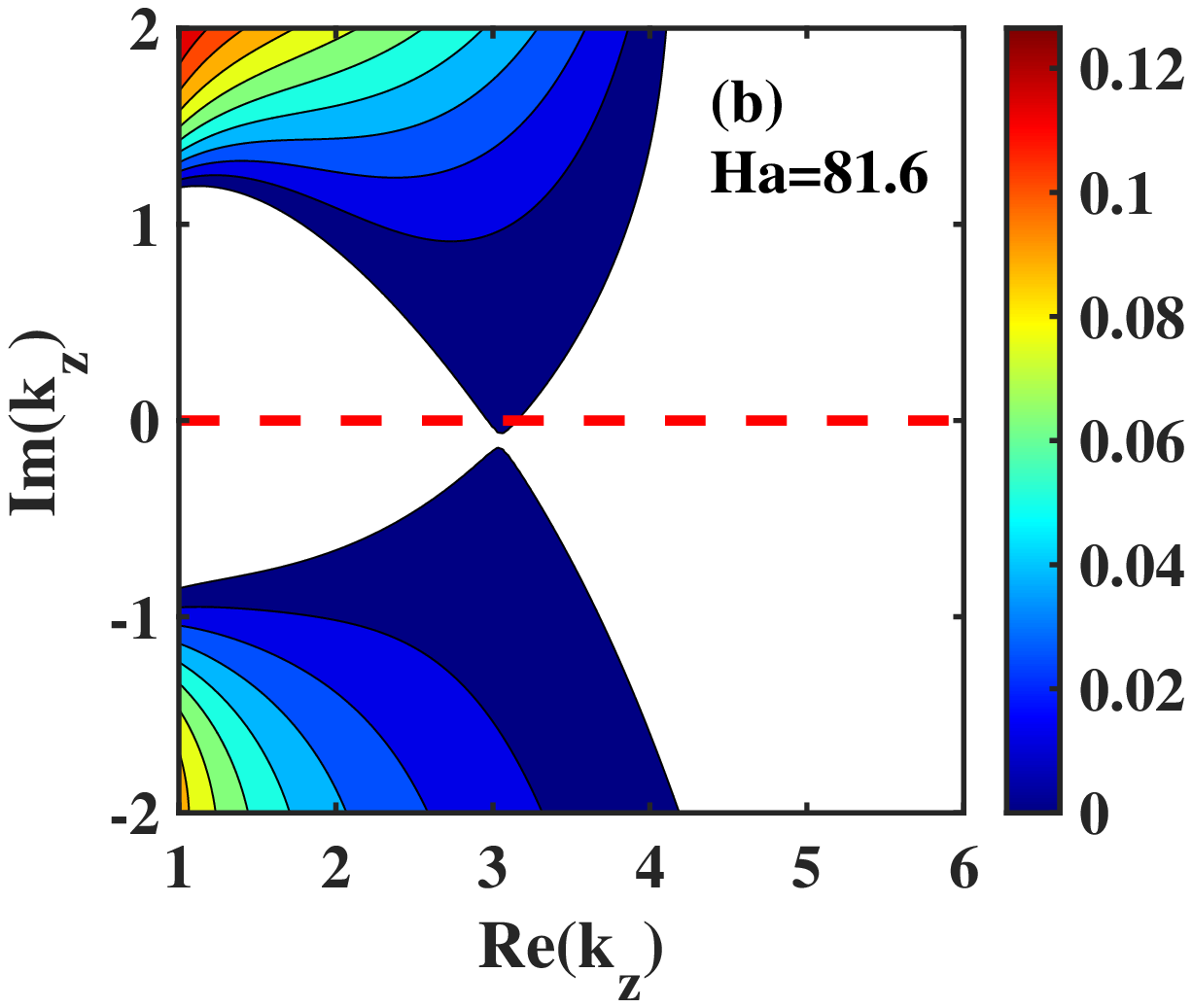}}
\end{minipage}
\begin{minipage}{.33\textwidth}
 \centerline{\includegraphics[width=\textwidth]{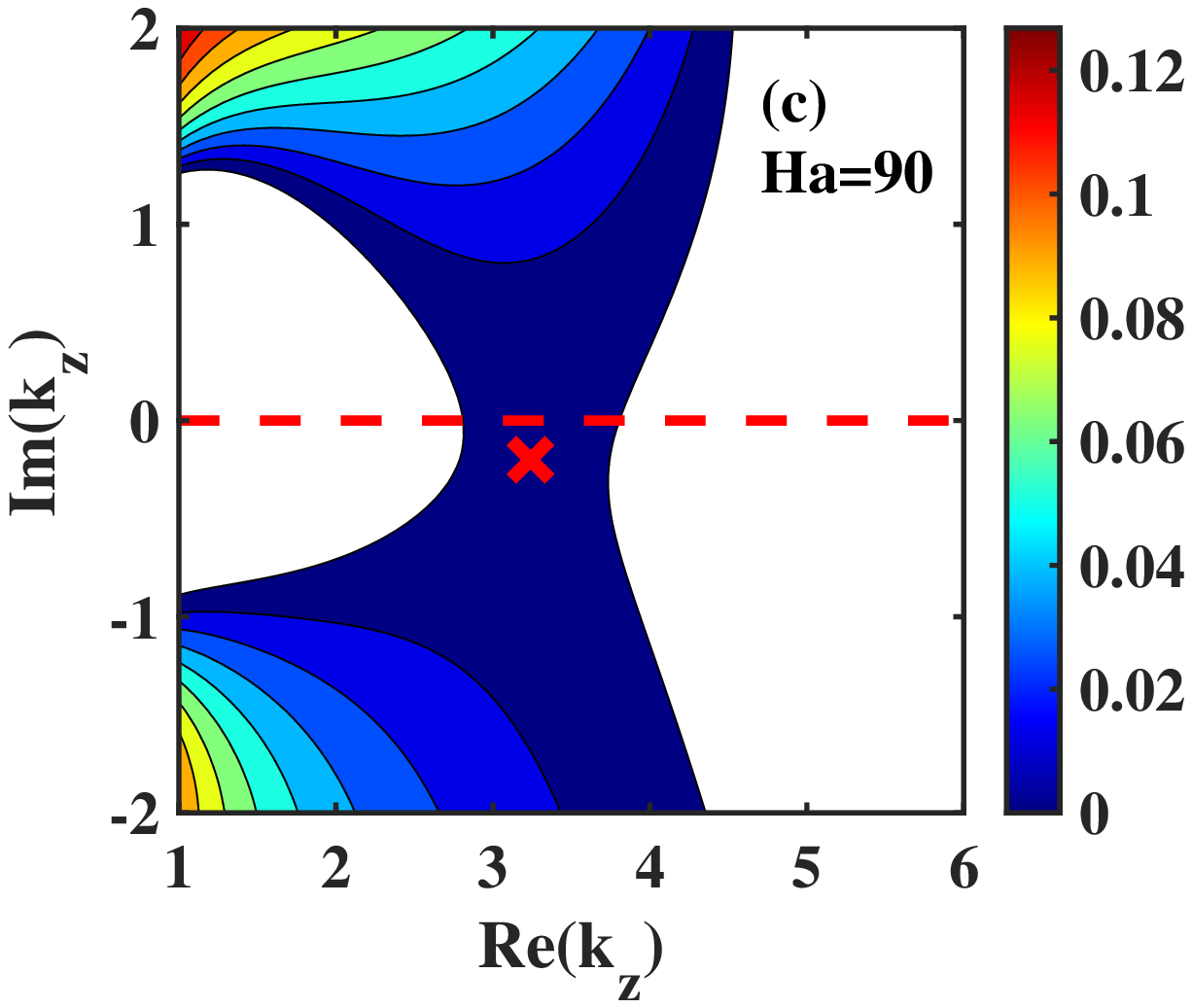}}
\end{minipage}
\begin{minipage}{.33\textwidth}
 \centerline{\includegraphics[width=\textwidth]{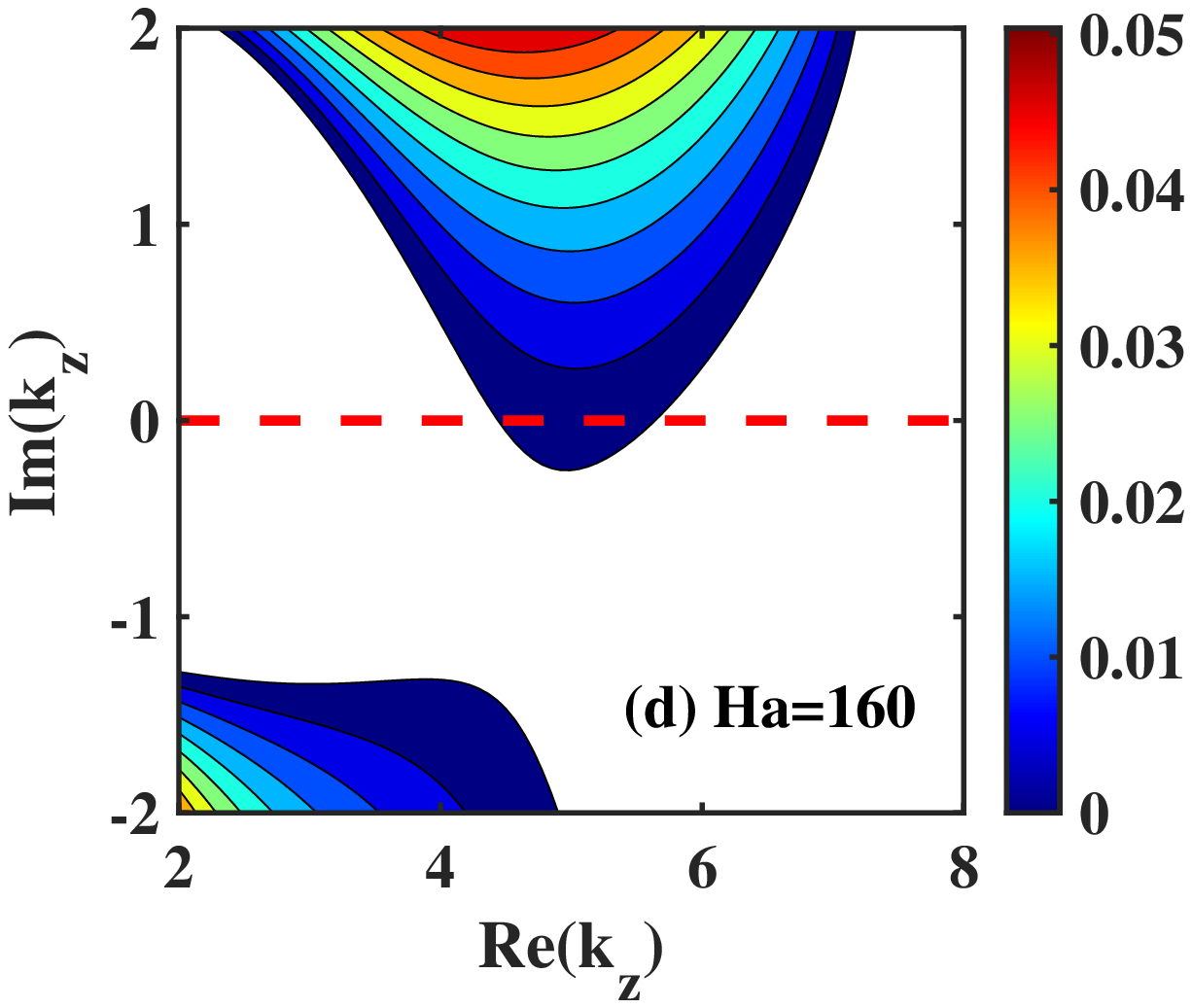}}
\end{minipage}
\begin{minipage}{.33\textwidth}
 \centerline{\includegraphics[width=\textwidth]{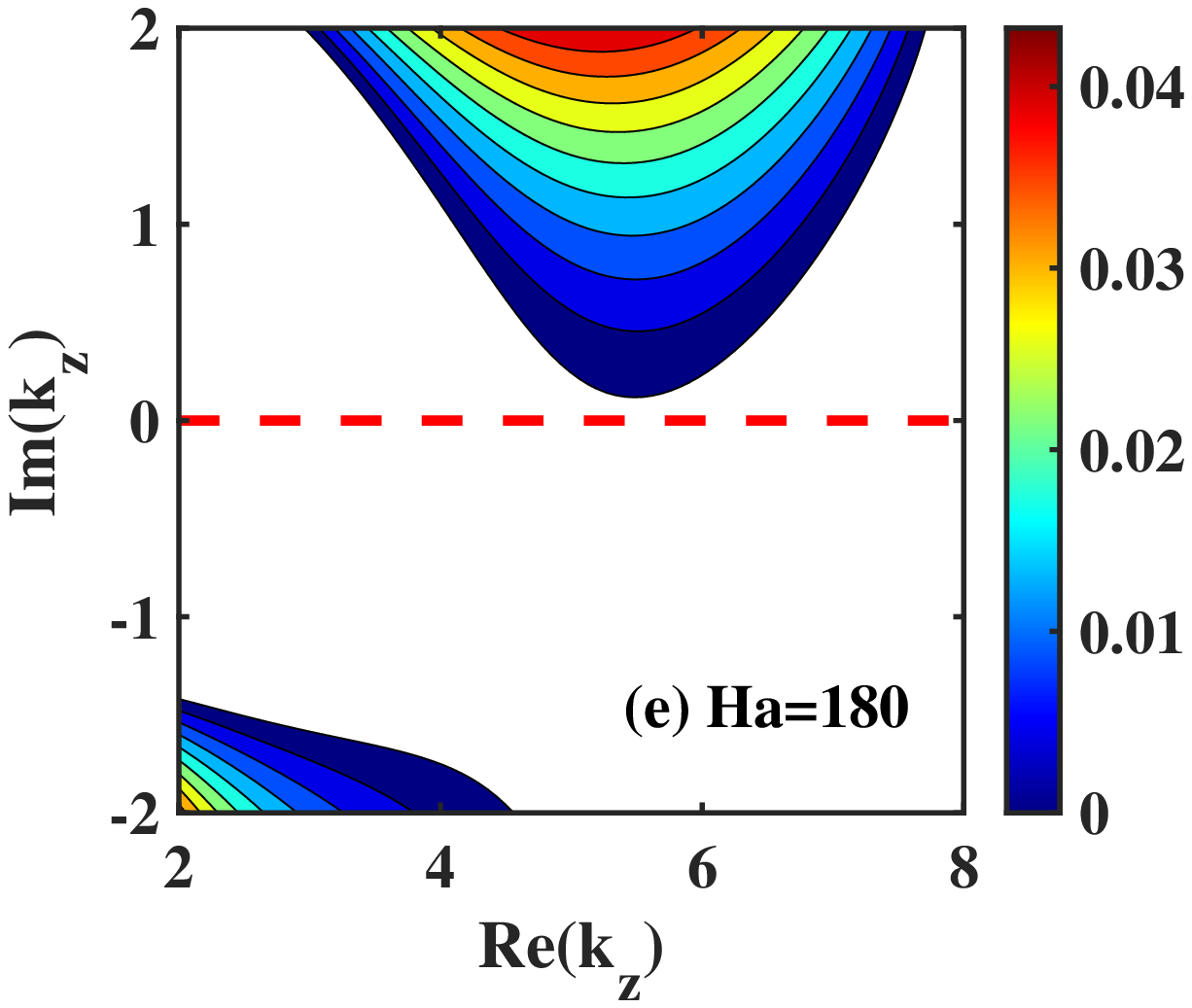}}
\end{minipage}
\begin{minipage}{.33\textwidth}
 \centerline{\includegraphics[width=0.85\textwidth]{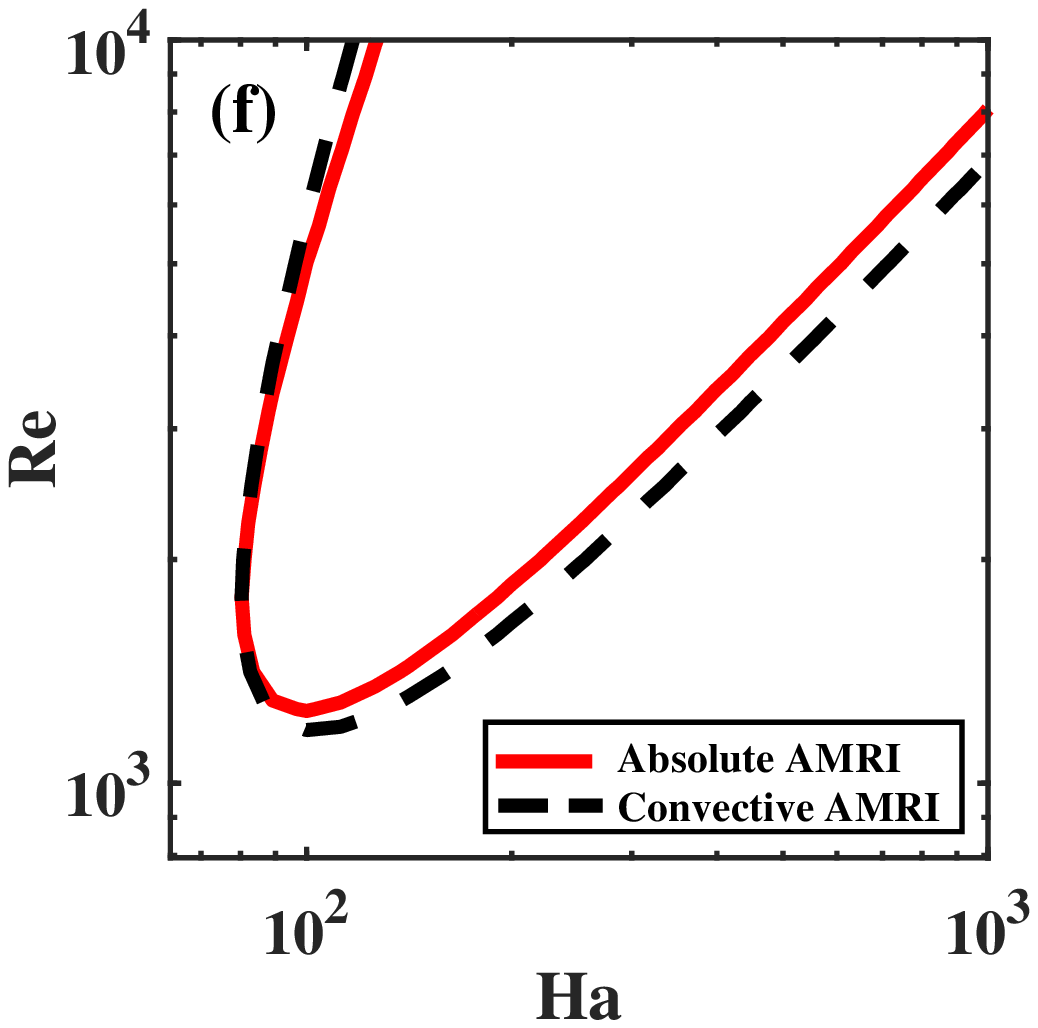}}
\end{minipage}
\caption{Same as in figure \ref{fig:WKB}, but for the 1D stability analysis at the same $Re=1480$ and $m=1$. Now the saddle point (red cross) in (\textit{c}), representing the absolute AMRI, is at $k_{z,s}=(3.24, -0.2)$ with the growth rate ${\rm Re}(\gamma(k_{z,s}))=1.9\times10^{-3}$.}\label{fig:1Danalysis}
\end{figure}
 
 \begin{figure}
 \begin{minipage}{.33\textwidth}
 \centerline{\includegraphics[width=\textwidth]{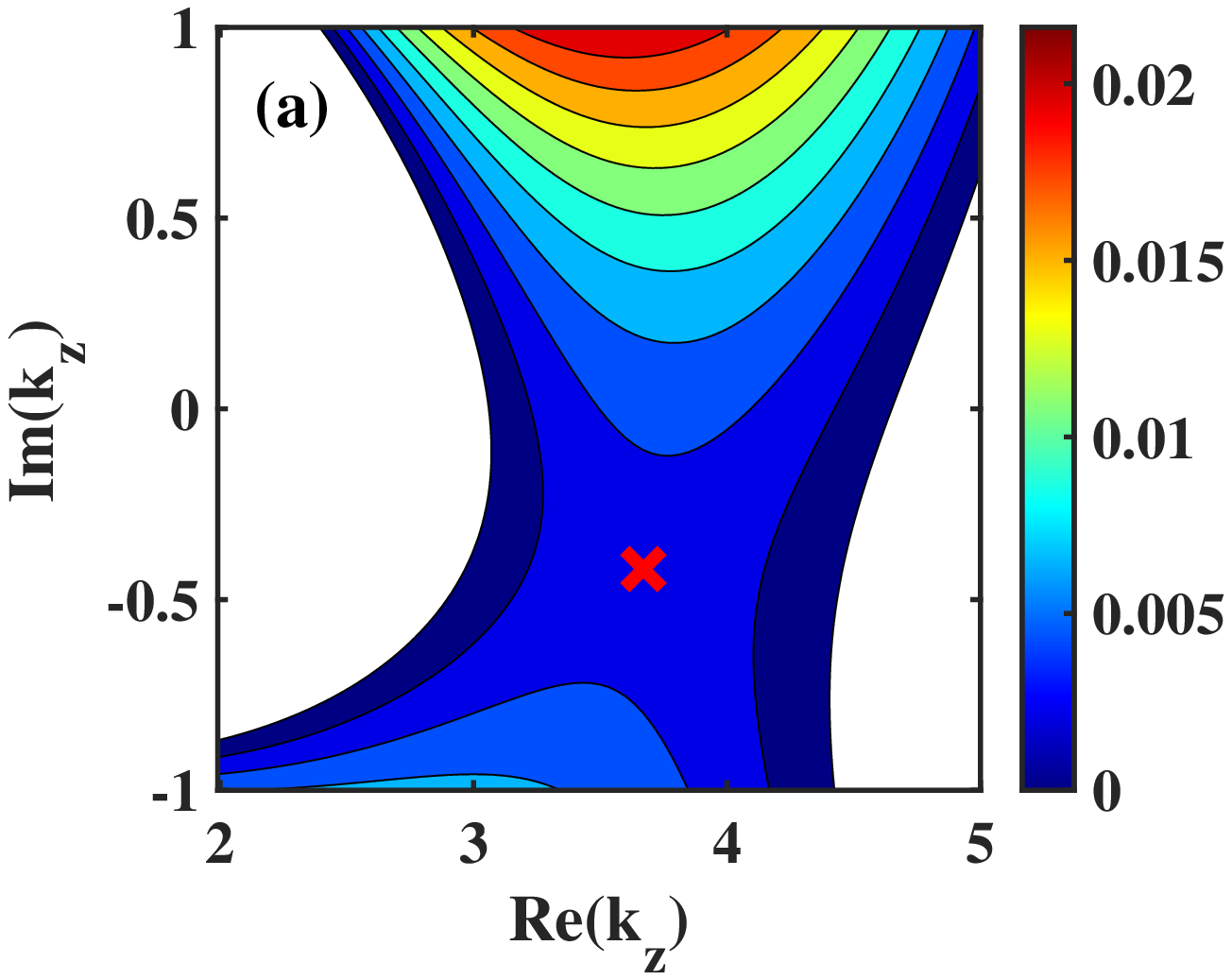}}
 \end{minipage}
 \begin{minipage}{.33\textwidth}
 \centerline{\includegraphics[width=\textwidth]{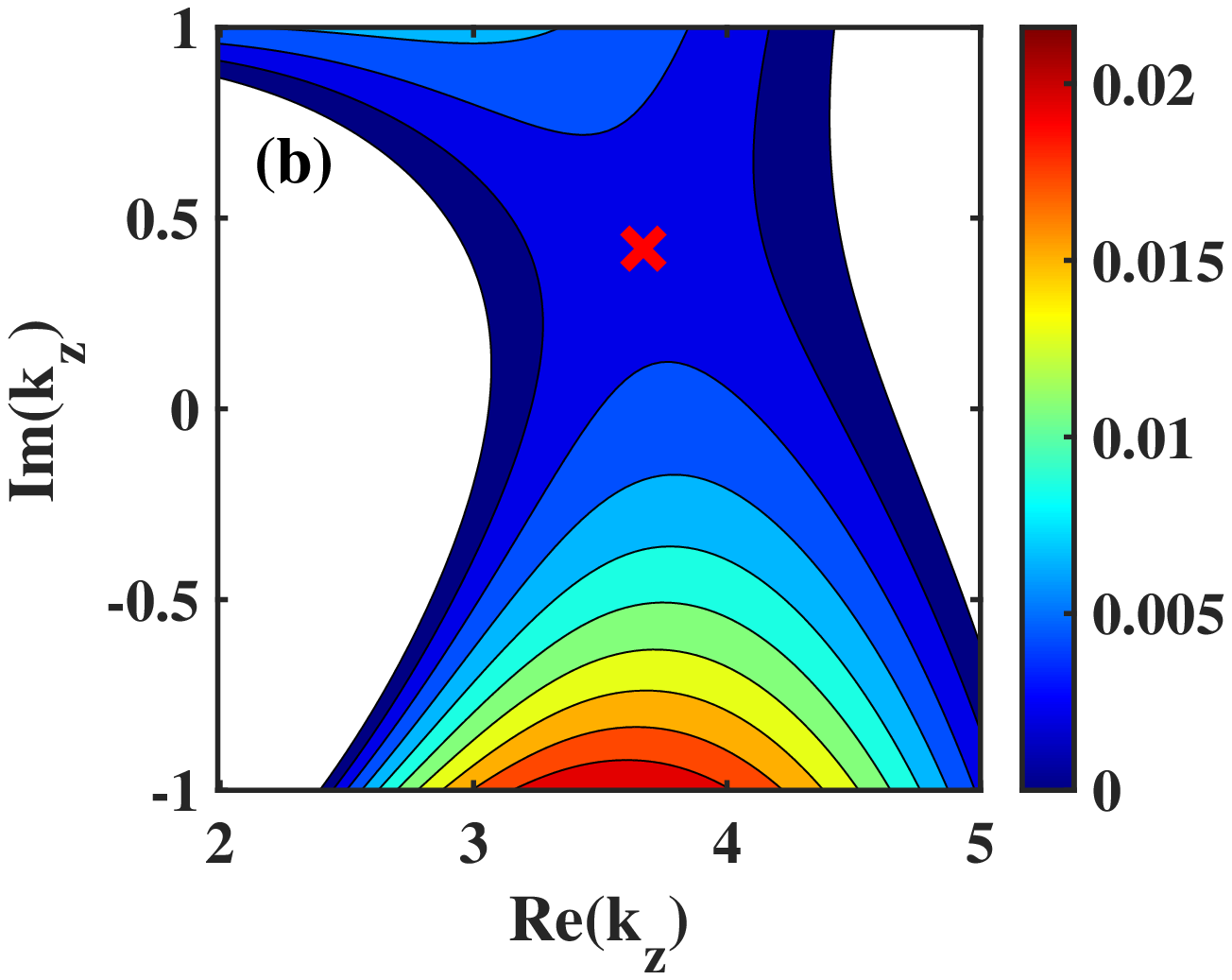}}
 \end{minipage}
 \begin{minipage}{.33\textwidth}
\centerline{\includegraphics[width=0.75\textwidth]{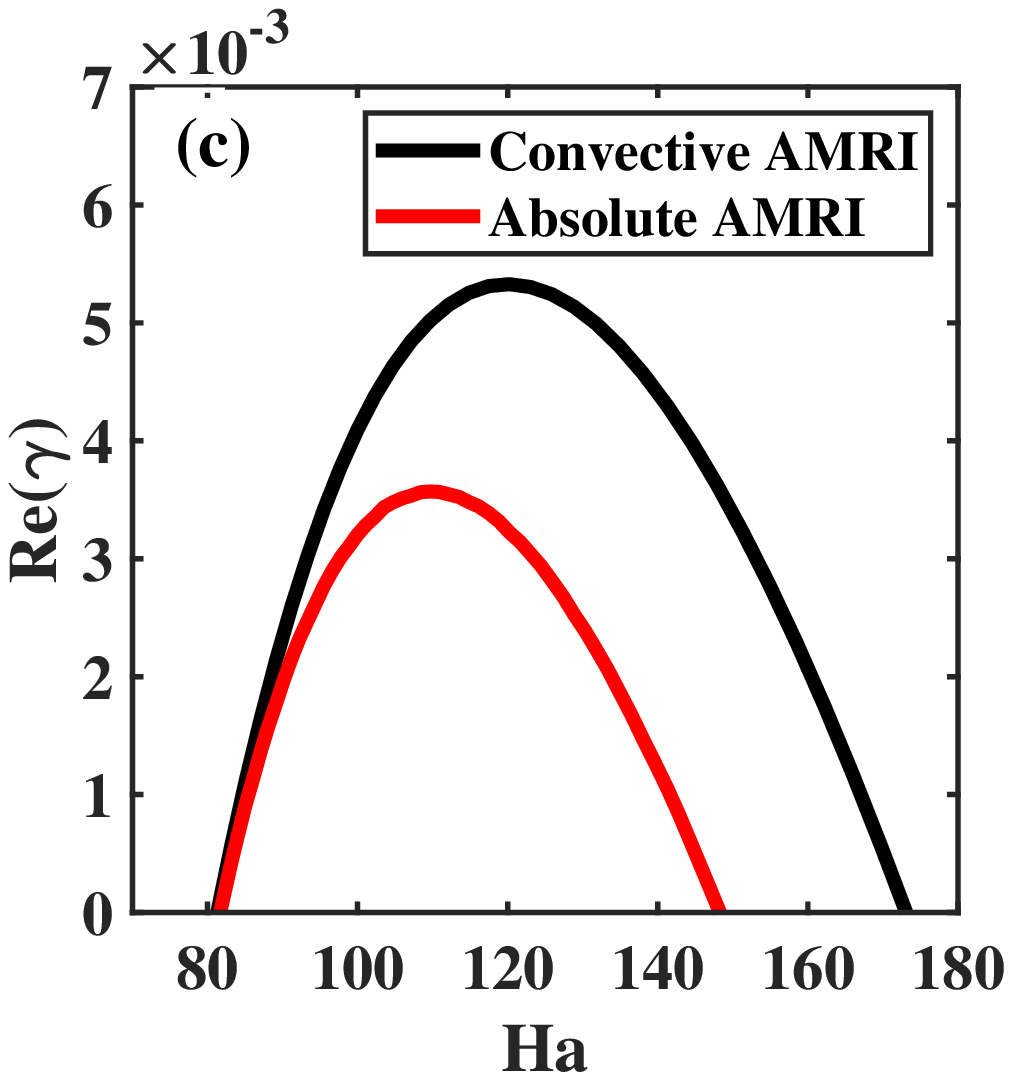}} \end{minipage}
\caption{The growth areas with ${\rm Re}(\gamma)>0$, in the $k_z$-plane at $Ha=110$ and $Re=1480$ in the 1D case separately for (\textit{a}) $m=1$ and (\textit{b}) $m=-1$ modes. Due to the mirror symmetry of the flow, these areas flip with respect to ${\rm Im}(k_z)$-axis when $m$ changes sign and, as a result, the saddle points (red crosses) for the absolute AMRI are symmetric around this axis: $k_{z,s}=(3.67, \mp 0.42)$ for $m=\pm 1$. The corresponding growth rate is ${\rm Re}(\gamma(k_{z,s}))=0.00356$ and the frequency is ${\rm Im}(\gamma(k_{z,s}))=\mp 0.2375$ for $m=\pm 1$. (\textit{c}) The growth rates of the convective (black) and absolute (red) AMRI vs. $Ha$ at the same $Re$.}\label{fig:m1_minus1_gr}
\end{figure}

\begin{figure}
\includegraphics[width=0.5\textwidth]{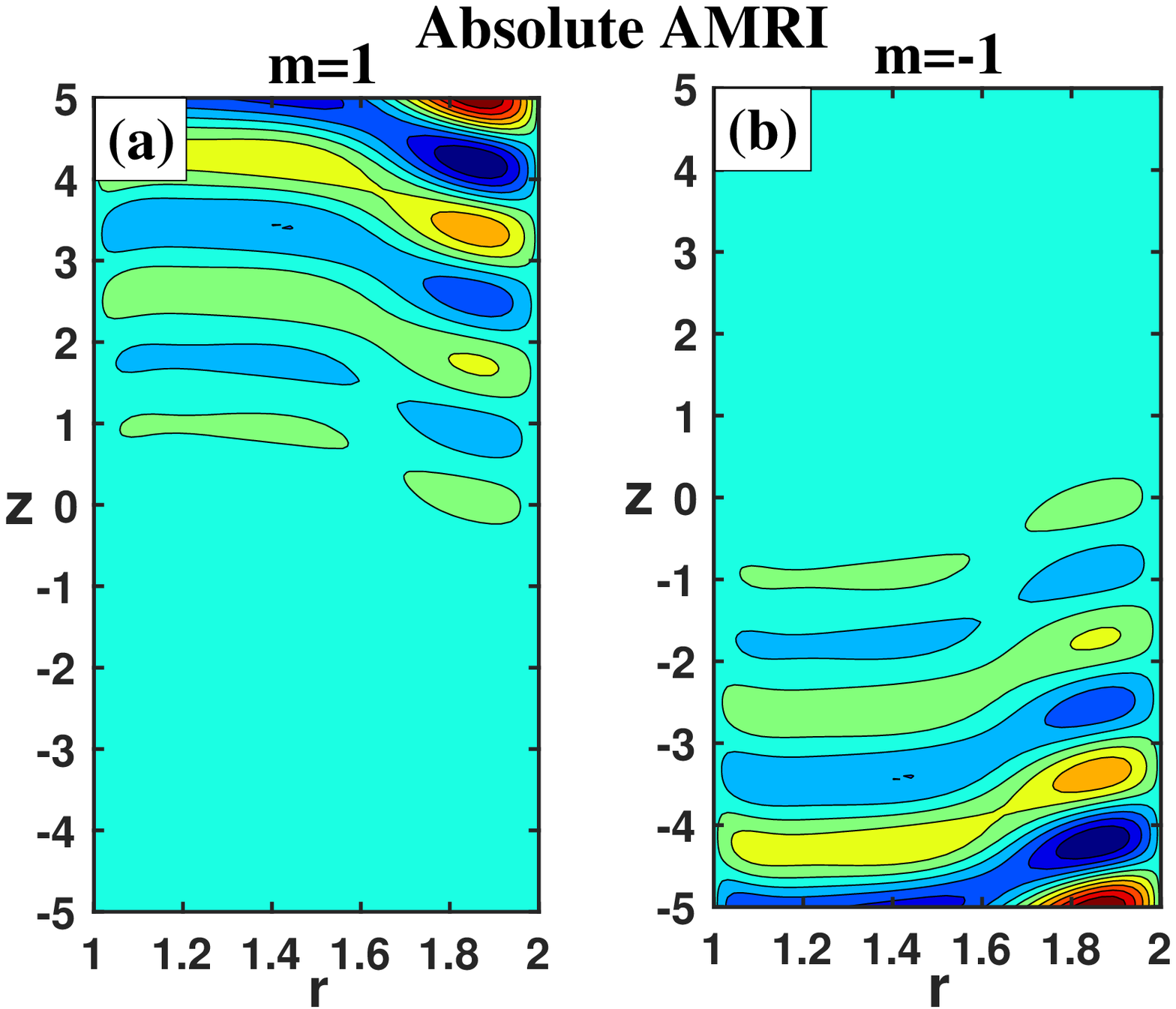}
\includegraphics[width=0.5\textwidth]{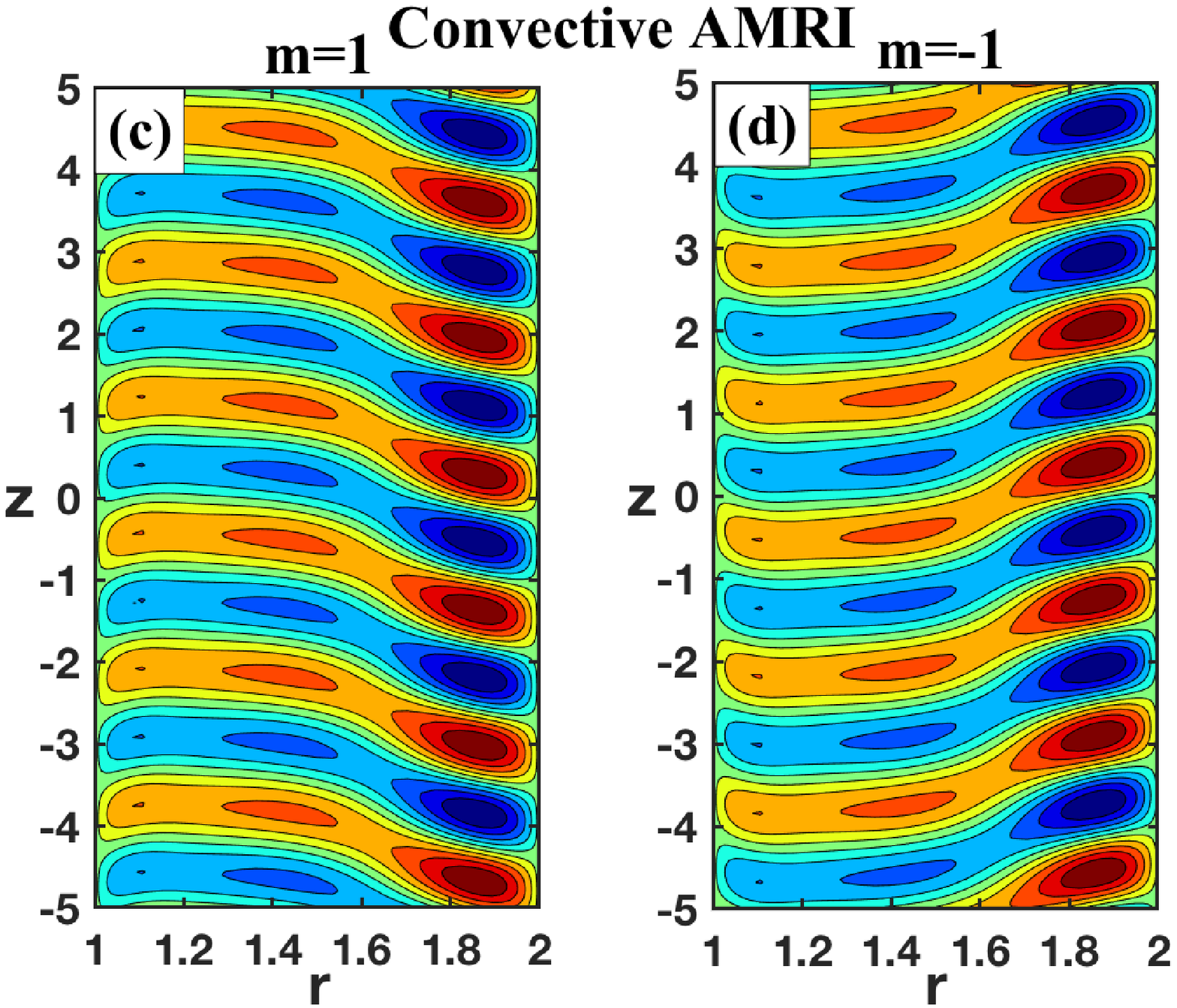}
\caption{Normalised axial velocity $u_z$ eigenfunctions in $(r,z)$-slice at $Ha=110$, $Re=1480$ pertaining to the absolute AMRI with (\textit{a}) $m=1$ and (\textit{b}) $m=-1$ (represented by red crosses in figure \ref{fig:m1_minus1_gr}) and to the convective AMRI with (\textit{c}) $m=1$ and (\textit{d}) $m=-1$.} \label{fig:eigenmodes}
\end{figure}


 \subsection {1D analysis }\label{sec:1D_analysis}
 
Now we present the solution of the 1D linear eigenvalue problem in the $k_z$-plane. The eigenvalues $\gamma$ and the radial structure of the associated eigenmodes are found using a spectral collocation method based on Chebyshev polynomials (up to $N=30-40$), whereby equations (\ref{eq:velocity})-(\ref{eq:zerodiv}), supplemented with the boundary conditions as described in Sec. \ref{sec:math}, are reduced to a large ($4N\times4N$) matrix eigenvalue problem \citep[see code details in][]{Hollerbach_etal2010}.
    
The resulting growth areas in the $k_z$-plane at the same $Re=1480$, $m=1$ and different $Ha$ are shown in figures \ref{fig:1Danalysis}(\textit{a})--\ref{fig:1Danalysis}(\textit{e}), which exhibit a similar behaviour as the WKB case above. Initially, for lower $Ha$, there is no AMRI, since both areas are distant from the ${\rm Re}(k_z)$-axis (red dashed line), figure \ref{fig:1Danalysis}(\textit{a}). When $Ha$ is increased to a critical value of 81.6, the convective AMRI sets in as the upper area crosses the ${\rm Re}(k_z)$-axis, figure \ref{fig:1Danalysis}(\textit{b}). This crossing point has a cusp-like shape and thus appears to be the saddle point at the same time, implying that the convective and absolute AMRI are nearly equivalent at the onset. The growth rate of the absolute AMRI further increases with $Ha$, as these areas merge. For example, at $Ha=90$ in figure \ref{fig:1Danalysis}(\textit{c}), the saddle point, corresponding to the absolute AMRI mode, is at $k_{z,s}=(3.24, -0.2)$ (red cross) with the growth rate ${\rm Re}(\gamma(k_{z,s}))=1.9\times10^{-3}$ and frequency ${\rm Im}(\gamma(k_{z,s}))=-0.25$. On further increasing $Ha$, first vanishes the absolute AMRI, figure \ref{fig:1Danalysis}(\textit{d}), and then the convective AMRI, figure \ref{fig:1Danalysis}(\textit{e}). 

In figure \ref{fig:1Danalysis}(\textit{f}), we also plot the marginal stability curves for the convective and absolute AMRI, with the latter being located inside the former, as in the WKB case, but in the 1D analysis, both are excited at almost similar critical $Ha$ and $Re$. Furthermore, comparing figures \ref{fig:WKB} and \ref{fig:1Danalysis}, shows that the distributions of the growth areas in the $k_z$-plane in the WKB and 1D analyses are qualitatively similar. However, the uncertainty in matching the constant $Ro$ in the WKB case and radially varying $Ro$ in the 1D case for a moderate gap width $(r_0-r_i)/r_i \sim 1$ \citep{Stefani_Kirillov2015} as well as some ambiguity in choosing the effective $k_r$ in the WKB analysis, makes quantitative differences between the two results unavoidable.

Figures \ref{fig:m1_minus1_gr}(\textit{a},\textit{b}) show areas with ${\rm Re}(\gamma)\geq 0$ in the $k_z$-plane at $Ha=110$ for which the absolute AMRI reaches a maximum growth for a given $Re=1480$. Since the TC flow with a purely azimuthal field is invariant to reversing the sign of $z$, the saddle points $k_{z,s}$ and hence the absolute AMRI at $m=\pm 1$ occur symmetrically with respect to the ${\rm Re}(k_z)$-axis: their ${\rm Re}(k_{z,s})=3.67$ and growth rate ${\rm Re}(\gamma(k_{z,s}))=0.00356$ are the same, while ${\rm Im}(k_{z,s})=\mp 0.42$ and frequency ${\rm Im}(\gamma(k_{z,s}))=\mp 0.2375$ differ only in sign. This implies that although the absolute AMRI modes have zero group velocity, they have non-zero phase velocities along and opposite the $z$-axis, respectively, for $m=-1$ and $m=1$. In figure \ref{fig:m1_minus1_gr}(\textit{c}), we also plot ${\rm Re}(\gamma)$ for the convective (optimized over ${\rm Re}(k_z)$) and absolute AMRI. It is seen that the convective AMRI has a larger growth rate and occurs for a wider range of $Ha$ than the absolute AMRI. As noted above, however, the convective and absolute AMRI are nearly identical at the onset for these parameters.

It is now interesting to look at the spatial structure of the absolute and convective AMRI modes. Figure \ref{fig:eigenmodes} shows the normalised eigenfunctions for the axial velocity $u_z$ in $(r,z)$-slice belonging to the absolute and convective AMRI at $m=\pm 1$, $Ha=110$ and $Re=1480$. It is evident that the eigenfunctions for these two types of AMRI are characteristically similar except that due to the complex wavenumber $k_{z,s}$ of the absolute AMRI, its eigenfunctions increase (for $m=1$) and decrease (for $m=-1$) along the $z$-axis. This property is important for the interpretation of the simulation results in Sec. \ref{sec:Simu}. By contrast, the wavenumber of the convective AMRI is real and hence its eigenfunctions retain a periodic structure with spatially constant amplitude along the $z$-axis. 

\section{Simulations}\label{sec:Simu}

We also conducted direct numerical simulation (DNS) of AMRI in the same magnetized TC flow setup and parameters as described in Sec. \ref{sec:math}, except that the cylinders now have a finite height $L=10r_i$ as in PROMISE. We  solved the basic equations of non-ideal MHD (in the same non-dimensional
form) using the open source code library OpenFOAM$^{\copyright}$\footnote{https://www.openfoam.com/}, complemented with a Poisson solver for the determination of the induced electric potential. The code used here has already been validated in our previous works \citep{Weber_2013,Seilmayer_etal2014}.
The initial conditions are the previous TC profile $\Omega(r)= C_1+C_2/r^2$ with the azimuthal magnetic field ${\bf B}_0=B_0(r_i/r){\bf e}_{\phi}$. For the velocity we set ${\bf u}=r\Omega(r){\bf e}_{\phi}$ (i.e., zero perturbation velocity) at the top and bottom of the cylinders and no-slip condition at the cylinder walls, while conducting boundary conditions are applied for the magnetic field, as in the 1D analysis above. Since the TC flow has a finite height and non-periodic boundary conditions along the $z$-axis, the AMRI developing in this case is in fact a {\it global} instability. For the same values of $Ha=110$ and $Re=1480$ as used in Sec. 4.2, we analysed the odd $m$ azimuthal modes of this global AMRI and compare it with the 1D linear analysis results. We note that although DNS solves full nonlinear equations, it is still in the linear regime, since the emerging global AMRI mode increases exponentially in time and has not yet reached nonlinear saturation (see below).

\begin{figure}
\begin{minipage}{0.52\textwidth}
\centerline{\includegraphics[width=\textwidth]{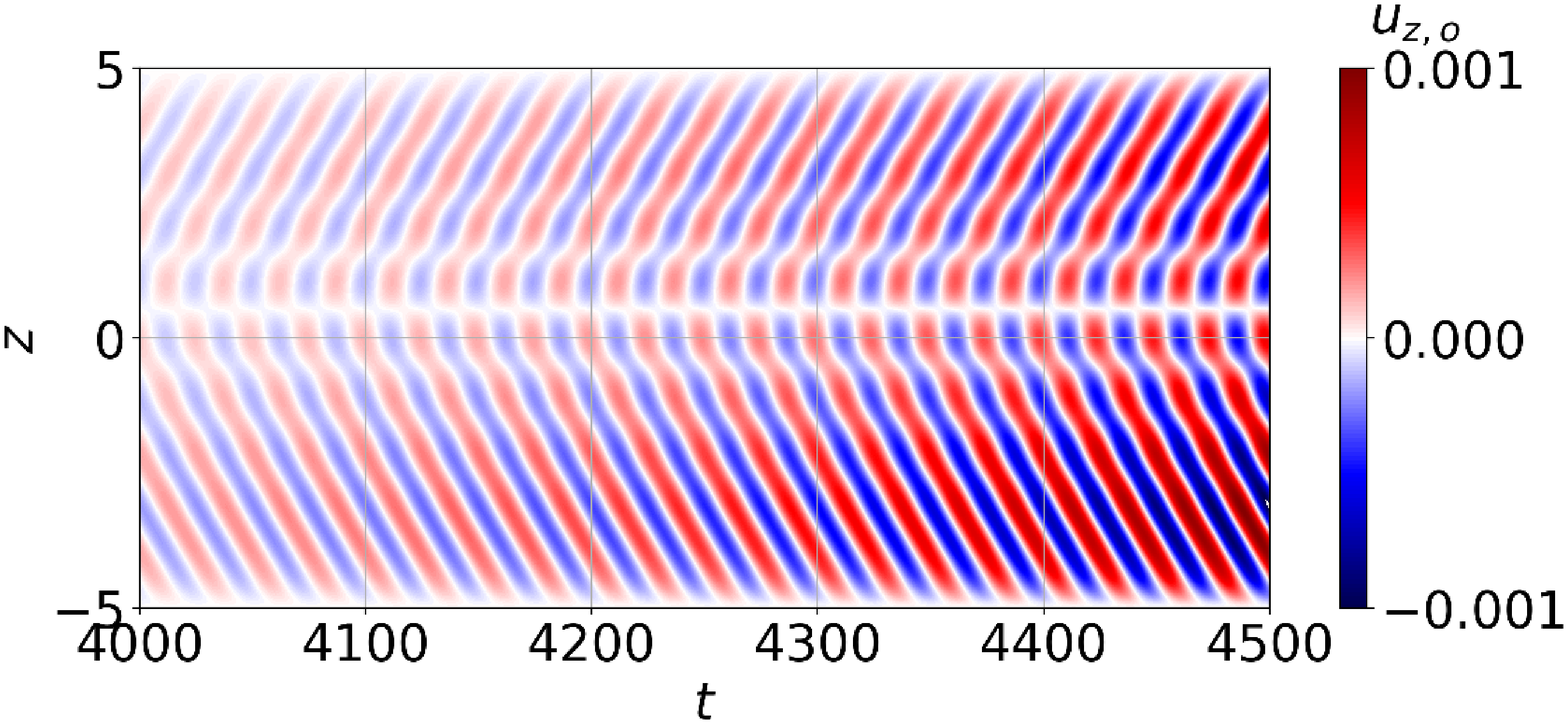}}
\end{minipage}
\begin{minipage}{0.48\textwidth}
\centerline{\includegraphics[width=\textwidth]{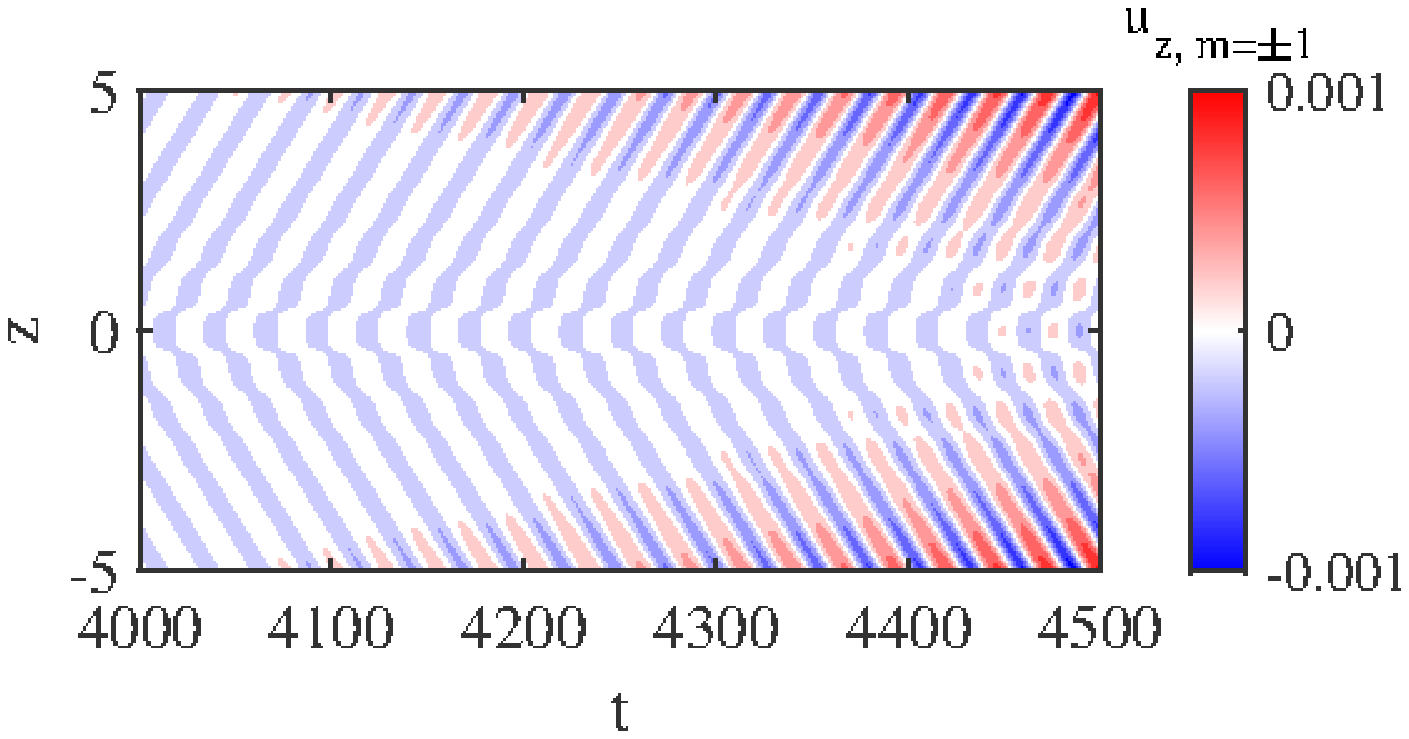}}
\end{minipage}
\setlength{\unitlength}{1mm}
\begin{picture}(1,1)
\put(8,27){(\textit{a})}
\put(79,27){(\textit{b})}
\end{picture}
\vspace*{-0.75em}
\caption{Spatio-temporal variation (butterfly diagram) of the axial velocity $u_z$ in $(t,z)$-slice at $Ha=110$ and $Re=1480$ for (\textit{a}) odd azimuthal modes of the global AMRI in the DNS and (\textit{b}) constructed from $m=\pm1$ eigenfunctions of the absolute AMRI in the 1D linear analysis, which are shown in figures \ref{fig:eigenmodes}(\textit{a,b}).} \label{fig:butterfly}
\end{figure}

In figure \ref{fig:butterfly}(\textit{a}), we show the temporal evolution of the $z$-profile of the axial velocity $u_{z,o}(z, t)$ at a radius $r_0=1.75$ for odd azimuthal modes of the global AMRI obtained from the DNS. To remove all even $m$ modes from the flow, we take the difference of the full axial velocity $u_z(r,\phi,z,t)$ at $\phi=0$ and $\phi=\pi$ for $r_0=1.75$, i.e., $u_{z,o}(z,t)=u_z(r_0,0,z,t)-u_z(r_0,\pi,z,t)$. Since $m=\pm 1$ are the most dominant non-axisymmetric modes in the flow, it is appropriate to compare this velocity profile with that of the absolute AMRI obtained by the 1D linear analysis. In figure \ref{fig:butterfly}(\textit{b}), we show the normalised axial velocity profile $u_z(z,t)$ for the absolute AMRI constructed from the linear superposition of its eigenfunctions at $m=\pm 1$ (figure \ref{fig:eigenmodes}\textit{a},\textit{b}), taken with equal weights. Figure \ref{fig:butterfly} shows a remarkable similarity in the butterfly patterns of the axial velocity resulting from the DNS and 1D analysis. Evidently, it also demonstrates  that the global AMRI in the simulations are dominated by the $m=\pm 1$ modes\footnote{We also checked this by directly calculating the amplitudes of different $m$ modes in the azimuthal Fourier transform of the global solution.} with the flow structure similar to that in the 1D analysis. Next, we obtain the frequency, growth rate and axial wavenumbers of this global AMRI structure. 

\begin{figure}
\begin{minipage}{0.48\textwidth}
\includegraphics[width=\textwidth]{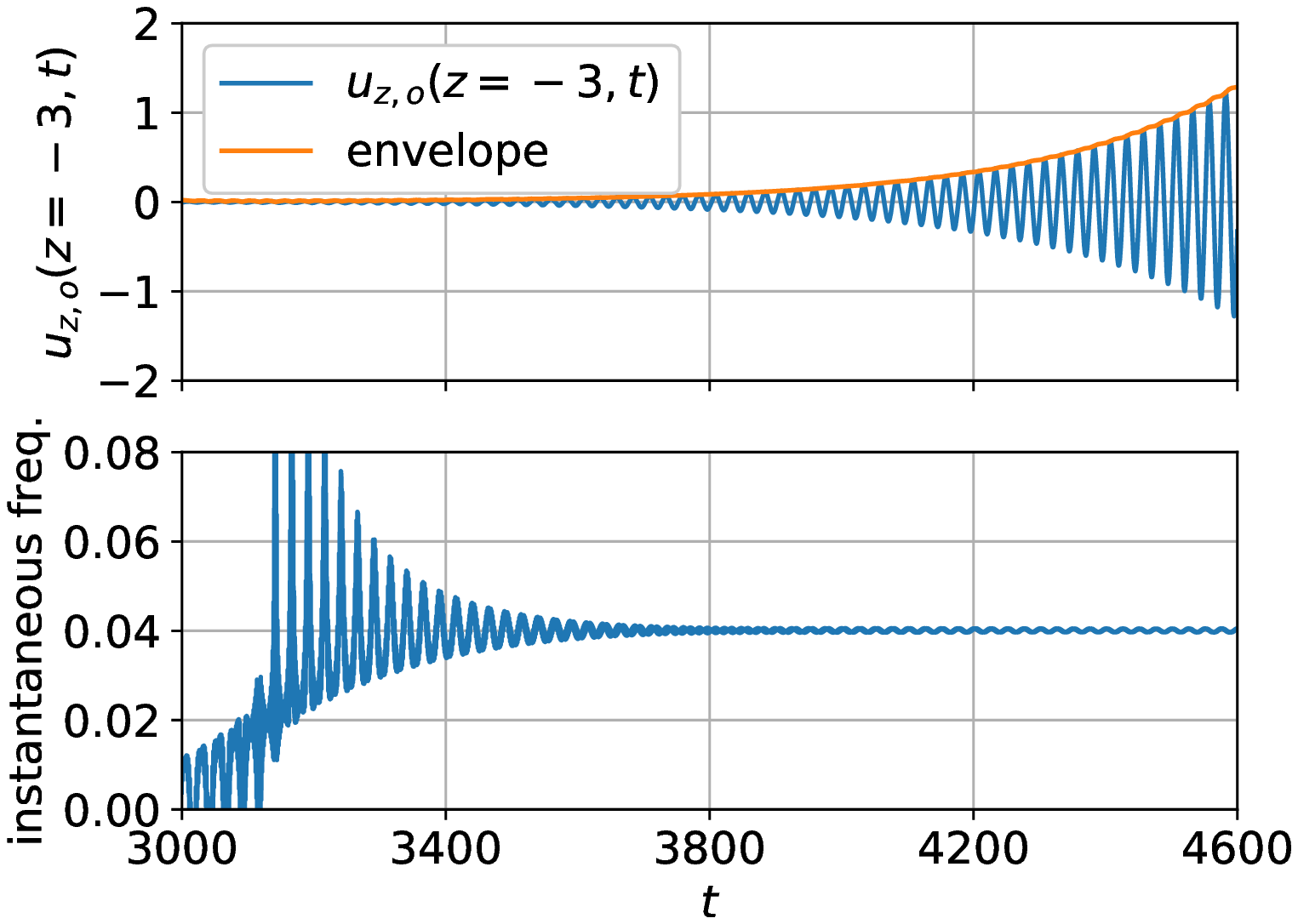}
\end{minipage}
\begin{minipage}{0.48\textwidth}
\includegraphics[width=1.05\textwidth]{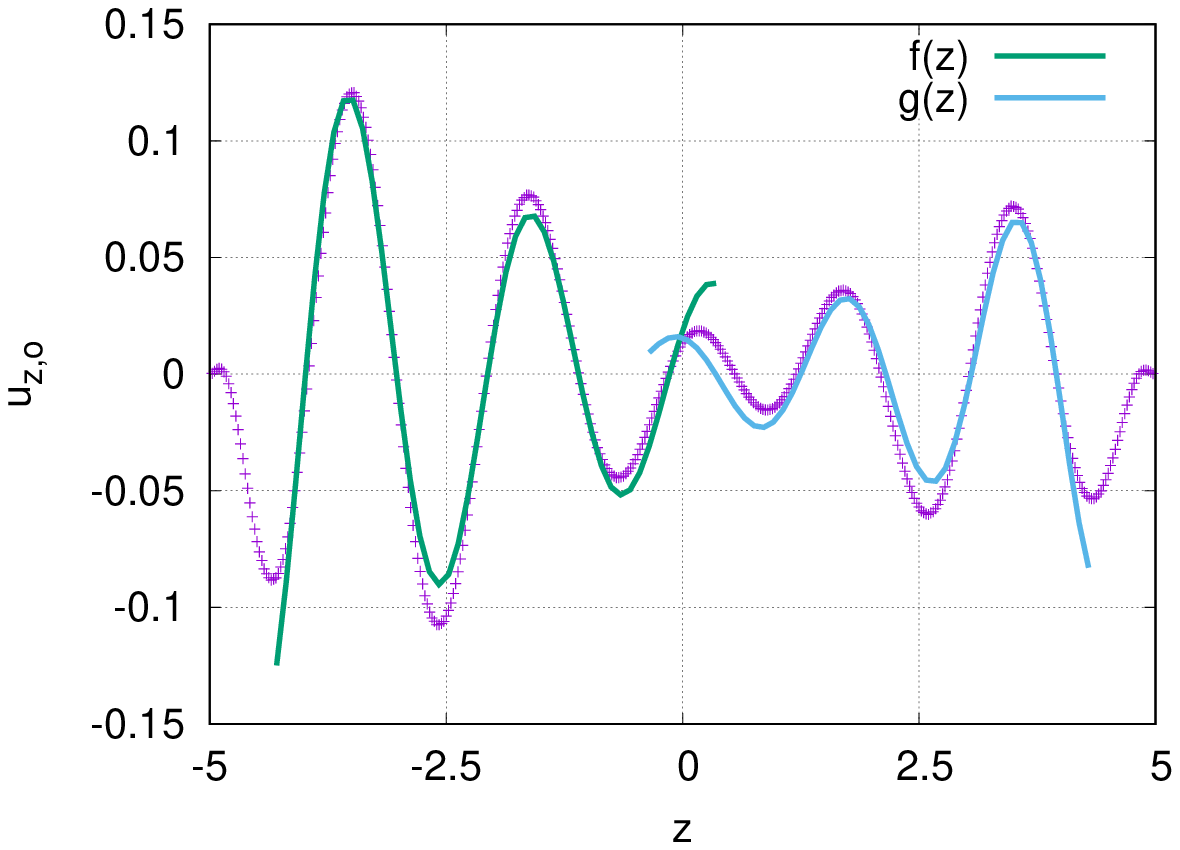}
\end{minipage}
\setlength{\unitlength}{1mm}
\begin{picture}(1,1)
\put(-79,19.5){(\textit{a})}
\put(-79,-1.5){(\textit{b})}
\put(-52,19.5){(\textit{c})}
\end{picture}
\vspace*{-0.75em}
\caption{(\textit{a}) Time evolution of $u_{z,o}(z_0,t)$ for the global AMRI at $z_0=-3$ with (\textit{b}) the corresponding Hilbert transform. (\textit{c}) Fits to the axial structure of this dominant global mode (lilac) obtained using POD at $r_0=1.75$ with two functions $f(z)$ (green) and $g(z)$ (blue), which correspond to the absolute AMRI modes with $m=-1$ and $m=1$, respectively (see text).}\label{fig:fitting}
\end{figure}

In figure \ref{fig:fitting}(\textit{a}), we plot $u_{z,o}$ at $z_0=-3$ as a function of time and make a fit with the expressions of the harmonic form
$A \exp(\Gamma t)\cos(\omega (t-t_0))$, obtaining in this way for the growth rate $\Gamma$ and frequency $\omega$ the values $\Gamma = 0.00337$ and $\omega=0.252$. We also conducted a Hilbert transform (figure \ref{fig:fitting}\textit{b}) to obtain envelopes of the velocity evolution and instantaneous frequency, which yields growth rate $0.0034$ and frequency $0.251$ of the global AMRI. Both of these results are in a good agreement with the 1D linear stability analysis for the absolute AMRI that yielded the growth rate $0.00356$ and frequency $0.2375$, figures \ref{fig:m1_minus1_gr}(\textit{a},\textit{b}).

We also performed Proper Orthogonal Decomposition (POD) of $u_{z,o}(z,t)$ in order to get the real and imaginary components of the axial wavenumbers $k_z$ of the global AMRI mode structure in figure \ref{fig:butterfly}(\textit{a}) using the model reduction library {\it modred} \citep{Belson_etal2014}. The obtained amplitude of this dominant global mode as a function of $z$ is plotted in figure \ref{fig:fitting}(\textit{c}). As the global AMRI flow actually consists of the two  $m=\pm1$ modes, each prominent on either side of the symmetry $z=0$-plane, it is reasonable to make two separate fits using the expressions $f(z)=A_1\exp(-{\rm Im}(k_{z,1}) (z+5))\cos({\rm Re}(k_{z,1})(z-z_1))$ and $g(z)=A_2\exp(-{\rm Im}(k_{z,2}) (z-5))\cos({\rm Re}(k_{z,2})(z-z_2))$, which are also shown in figure \ref{fig:fitting}(\textit{c}), each corresponding to either of the two absolute AMRI modes with $m=\pm 1$. Upon such a fitting, we obtain the values $k_{z,1}=(3.26, 0.29)$ and $k_{z,2}=(3.49, -0.392)$, which agrees reasonably well with our 1D analysis result $k_{z,s}=(3.67, \mp 0.42)$ from figures \ref{fig:m1_minus1_gr}(\textit{a},\textit{b}). However, it should be noted that the DNS are done for a finite length TC flow, while in the linear stability analysis, the cylinder height is assumed to be infinite. Hence, the minor quantitative difference between the DNS and the 1D linear analysis results may be attributable to the assumption of infinite length of the cylinders in the latter approach.

\section{Conclusion}\label{sec:conclu}

By performing the WKB and 1D linear stability analyses, we have shown that the absolute variant of AMRI corresponds to a saddle point in the complex axial wavenumber plane which, in physical space, appears as the characteristic solution exponentially increasing in the (negative or positive) axial direction. This is quite different from, and has a more experimental significance than, the usually considered convective AMRI with a real axial wavenumber. The corresponding global instability for the finite height TC flow (with non-periodic axial boundary conditions) was obtained by DNS, which has been demonstrated to represent the superposition of two absolute AMRI modes with $m=\pm 1$ from the 1D stability analysis of infinitely long cylinders. In other words, the butterfly diagram exhibited by the global AMRI mode can be understood as arising due to the presence of two dominant absolute AMRI wave modes with zero group velocities but with respective phase velocities of propagation directed upwards and downwards. These results further refine the interpretation of the butterfly-type structure observed in the PROMISE experiment \citep{Seilmayer_etal2014, Seilmayer_etal2020}, and will be essential for the design of new liquid sodium MRI experiment in the new DRESDYN project \citep{Stefani_etal2019}. Finally, we note that understanding the dynamics of AMRI is also important for establishing its relevance for the ``dead zones'' of accretion (protoplanetary) discs \citep{Gammie96, Lesur_etal14}, which are a cold, dense, very weakly ionized and therefore magnetically less active inner regions of discs, where high resistivity, resulting in $Pm \ll 1$, leads to the suppression of standard MRI.




\backsection[Acknowledgments]{This project has received funding from the European Union’s Horizon 2020 research and innovation programme under the ERC Advanced Grant Agreement No. 787544 and from the Shota Rustaveli National Science Foundation of Georgia (SRNSFG, Grant No. FR17-107). We thank R. Hollerbach for kindly providing the linear 1D code used here and J. Ogbonna for carefully reading the manuscript. We also thank the anonymous Reviewers for constructive reports that improved the presentation of our work.}

\backsection[Declaration of Interests]{The authors report no conflict of interest.}


\bibliographystyle{jfm}
\bibliography{jfm_biblio.bib}

\end{document}